\documentclass[aps,prx,reprint,showpacs,amsmath,amssymb,floatfix,superscriptaddress]{revtex4-2}

\usepackage{bm}

\usepackage{xcolor}
\usepackage{graphicx}
\usepackage{physics}
\usepackage[detect-weight=true, separate-uncertainty=true]{siunitx}
\usepackage{multirow}
\usepackage{enumitem}
\usepackage[version=4]{mhchem}

\usepackage[colorlinks,linkcolor=blue,urlcolor=blue,citecolor=blue,anchorcolor=blue]{hyperref}
\usepackage[capitalise]{cleveref}

\newcommand{\affIMNP}{CNRS, Aix-Marseille Universit\'{e}, Universit\'{e} de Toulon, IM2NP, Marseille, France.}
\newcommand{\affBIP}{CNRS, Aix-Marseille Universit\'{e}, BIP, Marseille, France.}
\newcommand{\affCEA}{Quantronics group, Service de Physique de l'\'Etat Condens\'e (CNRS, UMR\ 3680), IRAMIS, CEA-Saclay, Universit\'e Paris-Saclay, 91191 Gif-sur-Yvette, France}
\newcommand{\affIRCP}{Chimie ParisTech, PSL University, CNRS, Institut de Recherche de Chimie Paris, 75005 Paris, France}
\newcommand{\affPOMAM}{Institut de Chimie de Strasbourg, UMR 7177, CNRS-Universit\'e de Strasbourg, 67008 Strasbourg, France}

\begin{document}
	
	\title{Strongly anisotropic non-Kramers electron spin as a quantum coherence probe of angular fluctuations}

	\author{Achuthan~Manoj~Kumar}\affiliation{\affIMNP}
	\author{Remy~Dassonneville}\email{remy.dassonneville@im2np.fr}\affiliation{\affIMNP}
	\author{Guillaume~Gerbaud}\affiliation{\affBIP}
	\author{Nolwenn~Le~Breton}\affiliation{\affPOMAM}
	\author{Athanassios~K.~Boudalis}\affiliation{\affPOMAM}
	\author{Patrice~Bertet}\affiliation{\affCEA}
	\author{Philippe~Goldner}\affiliation{\affIRCP}
	\author{Sylvain~Bertaina}\email{sylvain.bertaina@cnrs.fr}\affiliation{\affIMNP}
	
	\date{\today}
	
	\begin{abstract}
		Strongly anisotropic non-Kramers rare-earth ions combine giant longitudinal $g$-factors with a vanishing transverse component imposed by time-reversal symmetry, a combination that makes their spin transitions exquisitely sensitive to the orientation of the applied magnetic field. We show that this sensitivity carries a dual identity: it is simultaneously an overlooked decoherence channel and the basis for a spin-coherence-based angular probe. Using pulsed electron paramagnetic resonance at X-band, we report the first measurements of the quantum coherence of \ce{Tb^3+} in a native-doped \ce{CaWO4} crystal (\SI{15}{ppb}) and map the Hahn-echo coherence time $T_2$ as a function of temperature (2 to \SI{10}{K}) and resonant field ($10^3$ to $10^4$~G). A parameter-free model combining spin-lattice relaxation, instantaneous diffusion and spectral diffusion from all independently quantified impurities overestimates $T_2$ by an order of magnitude at low temperature and wrongly predicts the field dependence of $T_2$, inconsistent with the observed monotonic decrease of $T_2$ with $B_r$. A two-parameter extension, including dynamical angular fluctuations of the crystal axis, reproduces the full dataset across multiple setups and laboratories. Two controlled experiments nominally identical except for different mechanical configuration of the setup establish the mechanical origin of the dominant contribution. 
        The two-parameter extension corresponds to an angular amplitude noise spectral density of overall order \SI{36}{\nano\degree/\sqrt{Hz}} from global external vibrations (ranging from 10 to \SI{66}{\nano\degree/\sqrt{Hz}} depending on the exact setup mechanical configuration) estimated at $\sim$\SI{2.5}{kHz} plus a temperature-dependent contribution assumed to come from local phonon-driven angular jitter. It identifies and highlights a decoherence pathway of practical relevance to any anisotropic solid-state spin system.
	\end{abstract}
	
	\maketitle
	
	\section{Introduction}
	Controlling and understanding decoherence is a central problem of quantum science. Electron spins in crystalline solids have emerged as a leading platform for quantum technologies, offering long coherence times, compatibility with microwave control, and rich coupling to optical, mechanical and nuclear degrees of freedom. Applications span microwave-to-optical quantum transduction~\cite{ZhongGoldner2019}, long-lived quantum memories~\cite{thiel2011,wolfowicz2021}, and quantum sensing~\cite{degen2017}. In this context, identifying and quantifying every relevant decoherence mechanism is a prerequisite for the rational engineering of quantum devices~\cite{LeDantec2021,rancic_electron_spin_2022}.
	
	Rare-earth ions with non-Kramers ground states occupy a peculiar corner of this landscape. Unlike Kramers ions, whose doubly degenerate ground states are protected by time-reversal symmetry, non-Kramers ions rely on the crystal field for any remaining degeneracy, which can be fully lifted by electric fields of sufficiently low symmetry. The resulting spin Hamiltonians feature symmetry-imposed vanishing transverse $g$-factors~\cite{griffith1963}, giant effective longitudinal $g$-factors, and a profound sensitivity of the transition frequency to the orientation between the crystal symmetry axis and the applied magnetic field~\cite{thiel2011,Kunkel_2018}. These properties make non-Kramers ions qualitatively distinct from their Kramers counterparts. Yet systematic studies of their quantum coherence in the pulsed-microwave regime remain scarce.
	
	A paradigmatic host for rare-earth spin studies is calcium tungstate, \ce{CaWO4}, a scheelite-structure crystal that has served as a reference matrix for EPR spectroscopy since the earliest days of the field~\cite{hempstead1960,forrester_paramagnetic_1962,mims_phase_1968}. Its low phonon density of states also makes it attractive for cryogenic detector applications such as direct dark matter search~\cite{MunsterPhD}. Renewed and sustained interest for \ce{CaWO4} in quantum technologies comes from its intrinsically low magnetic noise background: the only nuclear-spin-carrying tungsten isotope, \ce{^183W}, has a natural abundance of only \SI{14.3}{\percent}, and residual paramagnetic impurities introduced during crystal growth are typically at the parts-per-billion level~\cite{Billaud2025}. This quiet magnetic environment has enabled a sequence of landmark results on rare-earth spin coherence: spin-orbit qubit manipulation of \ce{Er^3+} with coherent Rabi oscillations~\cite{Bertaina2009}, a Hahn-echo coherence time of \SI{23}{ms} on dilute \ce{Er^3+} ensembles at millikelvin temperatures~\cite{LeDantec2021}, single-electron spin detection by microwave photon counting~\cite{wang2023,Billaud2025}, quantitative characterization of millikelvin spectral diffusion~\cite{rancic_electron_spin_2022}, week-long microwave spectral holes~\cite{Wang2025weeklong},  \ce{Er^3+} spin-photon entanglement \cite{Ourari2023, Uysal2025},
    sub-second spin and optical coherences for \ce{^171Yb^3+}~\cite{tiranov2026}, and individual nuclear qubits with second-scale coherences~\cite{OSullivan2025,Travesedo2025}. All of this work concerns Kramers ions. The present paper extends the scope of \ce{CaWO4} spin physics to its non-Kramers counterparts.
	
	For spins in anisotropic environments, the \emph{static} orientation of the crystal relative to the applied magnetic field is known to influence $T_2$: in silicon through the anisotropic dipolar coupling to the \ce{^29Si} nuclear bath~\cite{Tyryshkin_2006,Witzel2005}, in \ce{Er^3+}:\ce{Y2SiO5} through orientation-dependent erbium flip-flop rates~\cite{Bottger2009,Car2019}, in diamond NV centers through the anisotropic hyperfine coupling to \ce{^13C}~\cite{Stanwix2010}, and in hBN boron vacancies through analogous mechanisms~\cite{Mistri2025}. In all these cases the orientation \emph{selects} among different decoherence pathways but is itself a fixed parameter.
	
	A qualitatively distinct mechanism arises when the orientation \emph{fluctuates dynamically}. Any spin with finite $g$-factor anisotropy translates angular fluctuations into transition-frequency fluctuations, and thus into pure dephasing. External mechanical perturbations such as cryostat pulse tubes, turbo pumps, seismic activity, or acoustic room noise have been invoked as sources of degraded coherence in EPR~\cite{Kalra2016}, 
    in silicon-based spin qubits~\cite{Britton2016}, in isotropic spins through field-gradient coupling to sample motion~\cite{Ross2019}, in NMR~\cite{Allerhand,Sykora_fid,Sykora_hahn}, in optical coherences \cite{Louchet-Chauvet2019}, and in superconducting qubit platforms~\cite{Kono2024}. The reverse viewpoint, in which anisotropic spins serve as probes of mechanical fluctuations, has been demonstrated with NV centers coupled to nanomechanical oscillators~\cite{Arcizet2011,Kolkowitz2012,Ovartchaiyapong2014,Teissier2014} and, in a complementary optical-coherence setting, with rare-earth-doped crystals used as cryogenic vibration sensors~\cite{louchet_chauvet_2022}. The specific coupling pathway through \emph{angular} jitter in strongly anisotropic non-Kramers systems, however, has not previously been identified, modeled, or quantified.
	
	This work unifies the two viewpoints. We measure the Hahn-echo coherence of \ce{Tb^3+} in \ce{CaWO4} across one order of magnitude in resonant field and about a decade in temperature. A parameter-free standard model, built from independently measured spin-lattice relaxation, instantaneous diffusion, and spectral diffusion from every independently quantified impurity species, fails quantitatively in two distinct ways: it overestimates $T_2$ by an order of magnitude at low temperature, and it fails to reproduce the monotonic decrease of $T_2$ with resonant field $B_r$. A two-parameter extension based on angular jitter, coming from external mechanical vibrations and a phonon-driven local contribution, accounts for the complete dataset. Two controlled experiments, i ) at fixed temperature and field, switching a nearby mechanical pump on and off and ii) at fixed temperature, modifying the tightness of the screws supporting mechanically the cryostat insert, establish the mechanical origin of the dominant contribution. 
    
    The same coupling that makes \ce{Tb^3+} vulnerable to mechanical noise turns it into an angular probe with a few tens of nanodegree-per-square-root-Hertz-scale resolution in the kilohertz band. The mechanism identified here is not specific to \ce{Tb^3+}:\ce{CaWO4}; it applies to any anisotropic spin transition and should be considered in any precision coherence experiment involving Kramers and non-Kramers ions with mechanical elements in the vicinity of the sample.
	
	\section{Spin Hamiltonian and spectroscopic characterization}
	\label{sec:sample}
	
	\begin{figure}[!ht]
		\centering
		\includegraphics[width=\columnwidth]{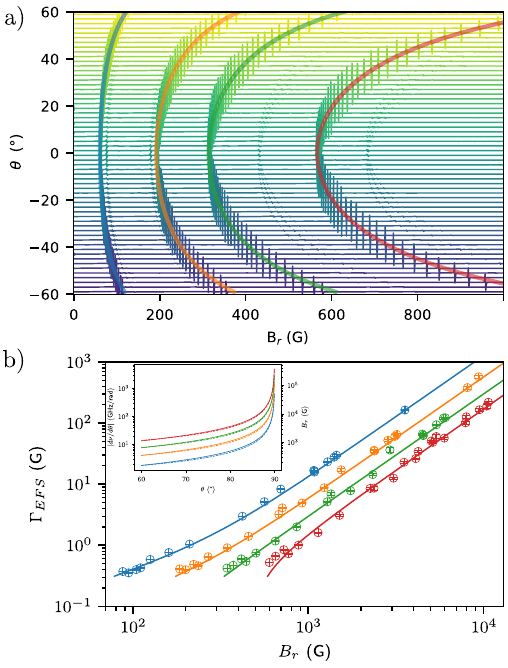}
		\caption{\textbf{a)} CW-EPR spectra at \SI{13}{K}, each offset vertically according to the orientation $\theta$ of the applied field relative to the crystal c-axis. The four hyperfine transitions of the effective spin-$1/2$ of \ce{Tb^3+} are labeled L1 to L4 (dashed colored lines), computed from \cref{eq:Bres} with $m_I = +1/2$ (L1, blue), $+3/2$ (L2, orange), $-1/2$ (L3, green) and $-3/2$ (L4, red). Near $\theta \sim 0$, all four lines fall below \SI{700}{G} owing to the large $g_\parallel$; they diverge as $\theta \to \pi/2$. Line L2 appears as a fold-back of a branch whose physical resonance occurs at negative applied field (see text). The other small visible lines are identified as \ce{Tb^3+} orthorhombic sites (see Appendix \ref{app:cw}). \textbf{b)} Half-maximum echo-field-sweep linewidth $\Gamma_\mathrm{EFS}$ extracted from Gaussian fits at \SI{7}{K} (same color code). Solid lines, model of \cref{eq:inhomogenuous} with $U_\Delta/\Delta = \SI{0.05}{\percent}$ and static mosaicity $U_\theta = \SI{0.05}{\degree}$. Crystal-field disorder dominates near $\theta \sim 0$; mosaicity governs the linewidth as $\theta \to \pi/2$ where the angular sensitivity diverges. Inset: computed angular sensitivity \cref{eq:sensitivity} (solid lines) and resonant field \cref{eq:Bres} (dash lines) as a function of $\theta$. Both diverge as $\theta \to \pi/2$.}
		\label{fig:EFS}
	\end{figure}
	
	\ce{CaWO4} is a scheelite-structure tetragonal crystal (space group $I4_1/a$, $a = b = \SI{0.524}{nm}$, $c = \SI{1.137}{nm}$) in which rare-earth ions substitute \ce{Ca^2+} sites of $S_4$ point symmetry~\cite{forrester_paramagnetic_1962,hempstead1960}. The crystal studied here was not intentionally doped during growth, yet retains paramagnetic impurities at the parts-per-billion to parts-per-million level (\cref{tab:concentration} and Appendix \ref{app:quantification}). Their role in decoherence is analyzed in \cref{sec:results}.

    \ce{Tb^3+} (4f$^8$, free-ion ground manifold $^7F_6$, $J = 6$) carries an
even number of 4f electrons: Kramers' theorem does not apply, and no
degeneracy of the crystal-field levels is protected by time-reversal
symmetry. In the $S_4$ crystal field of the \ce{Ca^2+} site, which couples
only states whose $J_z$ differ by a multiple of four~\cite{kirton1965}, the
two lowest eigenstates of the $^7F_6$ manifold are
$|\phi_1\rangle \simeq |J_z = +6\rangle$ and
$|\phi_2\rangle \simeq |J_z = -6\rangle$, each carrying a small admixture of
$|J_z = \mp 2\rangle$ (Appendix~\ref{app:ground}). The two states are
time-reversed partners; they would be degenerate if the crystal field did
not connect them. It does: both belong to the same singlet representation
of $S_4$, so a crystal-field matrix element couples $|\phi_1\rangle$ and
$|\phi_2\rangle$ directly and splits their symmetric and antisymmetric
combinations by $\Delta = \SI{8.136}{GHz}$ at zero
field~\cite{forrester_paramagnetic_1962}. The pair is thus a
\emph{quasi-doublet}: two nearby singlets mixed and split by the crystal
field alone, in contrast with a Kramers doublet whose degeneracy only a
magnetic field can lift. In the $\{|\phi_1\rangle, |\phi_2\rangle\}$ basis,
where the Zeeman and hyperfine interactions are diagonal, this
crystal-field coupling appears as the transverse term $\Delta\hat{S}_x$ of
an effective spin-$1/2$ coupled to the nuclear spin $I = 3/2$ of the
\SI{100}{\percent}-abundant \ce{^159Tb} isotope. The spin Hamiltonian
reads~\cite{forrester_paramagnetic_1962,griffith1963}:

	\begin{align}
		\label{eq:Hamiltonian}
		\hat{H}/h &= \frac{\mu_B}{h}\,\bm{B}\cdot\bm{g}\cdot\bm{S} + \bm{S}\cdot\bm{A}\cdot\bm{I} + \Delta_x\hat{S}_x + \Delta_y\hat{S}_y \notag \\
		&= \frac{\mu_B}{h}\cos\theta\,B\,g_\parallel\hat{S}_z + A_\parallel\hat{S}_z\hat{I}_z + \Delta\hat{S}_x,
	\end{align}
	with $h$ the Planck constant, $\mu_B$ the Bohr magneton, eigenvalues directly in frequency units, $g_\parallel = g_c = 17.777$, $A_\parallel = \SI{6.32}{GHz}$ and $\Delta = \sqrt{\Delta_x^2 + \Delta_y^2} = \SI{8.136}{GHz}$~\cite{forrester_paramagnetic_1962}. Time-reversal symmetry imposes a rigorous constraint on the $g$-tensor of any non-Kramers doublet: in its principal axes, two of the three diagonal components vanish identically~\cite{griffith1963}. In \ce{CaWO4}, the $S_4$ point symmetry aligns the principal axis of the doublet with the crystallographic c-axis, so $g_a = g_b \equiv 0$ exactly and we similarly take $A_\perp = 0$. The vanishing of $g_\perp$ has direct experimental consequences: it forbids EPR detection in the standard perpendicular geometry and requires a microwave field with a non-zero c-axis projection (Appendix \ref{app:pulseEPR}). Microscopic details on the quasi-doublet are collected in Appendices \ref{app:ground} and \ref{app:pulseEPR}.
	
	Treating the nuclear spin as quasi-static ($\hat{I}_z \to m_I$), the four transition frequencies of \ce{Tb^3+}, one for each $m_I \in \{-3/2,\,-1/2,\,+1/2,\,+3/2\}$, read
	\begin{equation}
		\label{eq:transition}
		\nu = \sqrt{\!\left(\frac{\mu_B}{h}\,g_\parallel B\cos\theta + A_\parallel m_I\right)^{\!2} + \Delta^2}.
	\end{equation}
	At the cavity resonance $\nu_c$ ($\sim \SI{9.6}{GHz}$ at X-band), the corresponding resonant fields are
	\begin{equation}
		\label{eq:Bres}
		B_r(\theta, m_I) = \frac{h}{\mu_B}\,\frac{\sqrt{\nu_c^2 - \Delta^2} - A_\parallel m_I}{g_\parallel\cos\theta}.
	\end{equation}
	\cref{eq:Bres} reveals that $B_r$ diverges as $\theta \to \pi/2$, reflecting the vanishing effective $g$-factor $g_\parallel\cos\theta$. While $B_r$ diverges, the field amplitude projected along the c-axis $B_r \cos(\theta)$ remains constant. 
	
	Because $g_\perp = 0$, detection of the \ce{Tb^3+} EPR signal requires a non-zero projection of both the static and the microwave magnetic fields along the c-axis (Appendices \ref{app:clock_transition} and \ref{app:pulseEPR}). 
    Resonant fields as a function of orientation were measured with a parallel-mode cavity (\cref{fig:EFS}a). The four lines L1 to L4, indexed by increasing resonant field and corresponding to $m_I = +1/2,\,+3/2,\,-1/2,\,-3/2$, are well reproduced by \cref{eq:Bres} across the full angular range. For $m_I = +3/2$, we have from \cref{eq:Bres} $B_r<0$. This line at negative field is however folded-back and observed at positive $|B_r|$ (\cref{fig:EFS}.a).  

Using echo-field-sweep (EFS), we extract by Gaussian fits the half-maximum inhomogeneous linewidth $\Gamma_\mathrm{EFS}(B_r)$ for the four terbium lines (\cref{fig:EFS}b). The linewidths evolve from $<\SI{0.5}{G}$ at $\theta \sim 0$ to above \SI{100}{G} as $\theta \to \pi/2$. Inhomogeneous broadening arises from the static distribution of local environments, including growth-related strains, dislocations, and lattice imperfections~\cite{Schweiger2001PulsedEPR,AbragamBleaney1970}. Two contributions dominate: a static mosaicity, the angular spread of crystallite orientations within the macroscopic sample, and a distribution of crystal-field parameters arising from local strains. We model the inhomogeneous linewidth as~\cite{kirkby_epr_1967}
	\begin{equation}
		\label{eq:inhomogenuous}
		\Gamma_\mathrm{inh}^2 = \left(U_\Delta\,\pdv{\nu}{\Delta}\right)^{\!2} + \left(U_\theta\,\pdv{\nu}{\theta}\right)^{\!2}.
	\end{equation}
    To compare to the data, the inhomogeneous linewidth $\Gamma_\mathrm{inh}$ in \si{rad/s} is translated into magnetic field units using the computed derivative $\vert \dv{(2\pi\nu)}{B}\vert^{-1}$.
	A relative crystal-field variation $U_\Delta/\Delta = \SI{0.05}{\percent}$ combined with a static mosaicity $U_\theta = \SI{0.05}{\degree}$ reproduces the observed EFS linewidths across all four terbium lines (\cref{fig:EFS}b). The extracted mosaicity is consistent with X-ray rocking-curve measurements on \ce{CaWO4} crystals of comparable structural quality~\cite{MunsterPhD}. Crystal-field disorder dominates near $\theta \sim 0$, while mosaicity governs the linewidth as $\theta \to \pi/2$.
    
	Two derived quantities set the stage for the decoherence analysis below. 
    Due to $g_\perp = 0$, the magnetic dipole-dipole interaction only couples the \ce{Tb^3+} spin through the c-axis terms (see Appendix \ref{app:model}). For a given magnetic field $\bm{B}$, in general, the terbium magnetic moment, the spin bath magnetic moment and the magnetic field are not collinear. The dipolar linewidth from spin $b$ in the bath is given by (see Appendix \ref{app:model}):
    \begin{equation}
        \Gamma_{dd}(\theta, \mathrm{Tb}, \mathrm{b}) = \frac{\mu_0 C_b}{3 \hbar} \abs{\bm{\mu_\mathrm{Tb}}} \abs{\bm{\mu_{b}}(\theta) } \mathcal{J}(\psi(\theta, \mathrm{Tb}, \mathrm{b}))
        \label{eq:Gamma_dd}
    \end{equation}
    with $C_b$ the concentration of spin $b$ and $\mathcal{J}$ a geometric factor close to unity, that depends on the angle $\psi(\theta, \mathrm{Tb}, \mathrm{b})$ between the two magnetic moment $\bm{\mu_\mathrm{Tb}}$ and $\bm{\mu_{b}}$. Importantly, the magnitude of the terbium magnetic moment is given by $\abs{\bm{\mu_{Tb}}} = \mu_B g_\parallel \sqrt{\nu_c^2 - \Delta^2}/\nu_c$ and is independent of magnetic field magnitude $\abs{\bm{B}}$ and orientation $\theta$ at its resonant field $\bm{B_r}(\theta, m_I)$. The angular dependency of the dipolar linewidth is then mostly given by $|\bm{\mu_b}(\theta)|$.
    
    The second quantity is the angular sensitivity of the transition frequency given by
	\begin{equation}
		\label{eq:sensitivity}
		\left.\pdv{\nu}{\theta}\right|_{\bm{B_r}} = -\frac{\sqrt{\nu_c^2 - \Delta^2}}{\nu_c}\cdot\frac{\mu_B\,g_\parallel\,B_r\sin\theta}{h},
	\end{equation}
	which vanishes at $\theta = 0$ and diverges as $\theta \to \pi/2$. The sensitivity as well as the resonant field are plotted as a function of orientation $\theta$ in the inset of \cref{fig:EFS}.b.  
    
    Both expressions \cref{eq:sensitivity,eq:Gamma_dd} share a common suppression factor $\sqrt{\nu_c^2 - \Delta^2}/\nu_c$, which vanishes at the clock transition ($\nu_c = \Delta$; see Appendix \ref{app:clock_transition}) and tends to unity far from it. The limits of these equations are worth stressing.
     First, the magnetic moment of terbium is always along the c axis. The dipolar linewidth from terbium spin (from the same probed line or from an unprobed line) is then constant and does not depend on the field orientation $\theta$. It would also be the case for any other non-Kramers ions (which we don't have in our sample), as their transverse g-factor is also set to zero. For a Kramers ion $k$, the magnitude of the magnetic moment  is given by $|\bm{\mu_k}(\theta)| = \mu_B \frac{\sqrt{g_{\perp k}^4 \sin^2\theta + g_{\parallel k}^4 \cos^2\theta}}{g_k(\theta)} \ge \mu_Bg_k(\theta)$ with $g_k(\theta)$ its effective g-factor setting its transition frequency. 
     Thus, the dipolar linewidth decreases with $\theta$ for Kramers ions with $g_{\parallel, k} > g_{\perp, k}$ (in our case, \ce{Ce^3+}, \ce{Dy^3+} and \ce{Mo^5+}), and reciprocally increases with $\theta$ if $g_{\parallel, k} < g_{\perp, k}$ (in our case, \ce{Er^3+} and \ce{Nd^3+}).
     However, the dipolar linewidth is not the only figure of merit setting the spectral diffusion. It also depends on the spin bath polarization and the rate $R$ at which the spin in the bath changes state. Due to $g_\perp = 0$, as $\theta \to \pi/2$, $B_r \to +\infty$, the transition frequency $\nu_k$ of the Kramers spin also tends to $+\infty$. At a given temperature $T$, they get more and more polarized, thus reducing the spectral diffusion. Moreover, the spin-lattice direct process increases as $\nu^5\coth(h\nu/2k_B T)$~\cite{AbragamBleaney1970}. At some point, the spectral diffusion enters the regime $R\tau \gg 1$ of fast motional-narrowing \cite{,bottger_optical_2006,HuHartmann1974} where the resulting effective $T_2$ scales with $R$, thus also suppressing the spectral diffusion (see Appendices \ref{app:T1_field} and \ref{app:SD}). 
Conversely, $\partial\nu/\partial\theta$ diverges in the same limit $\theta \to \pi/2$ (see \cref{fig:EFS}b inset), so angular fluctuations become the 
     \emph{main} decoherence channel. This opposing scaling is the mathematical backbone of the sensing modality developed in \cref{sec:sensing}.	
     
     A key distinction must be drawn at this point. The extracted mosaicity $U_\theta = \SI{0.05}{\degree}$ (\cref{fig:EFS}.b) is a \emph{static} angular spread over the ensemble of spins, refocused by the Hahn-echo $\pi$ pulse and therefore unrelated to $T_2$. The angular jitter identified below 
     is a \emph{dynamical} fluctuation of the crystal c-axis 
     relative to the fixed laboratory field direction. The two effects are physically distinct (ensemble disorder versus temporal noise) and differ by more than four orders of magnitude in amplitude. Accessing the latter requires both a spin-echo protocol and a complete model of all decoherence contributions, to which we now turn.
	
	\section{Results}
	\label{sec:results}
	
	\subsection{Spin-lattice relaxation}
	\label{sec:T1}
	
	\begin{figure}[!ht]
		\centering
\includegraphics[width=\columnwidth]{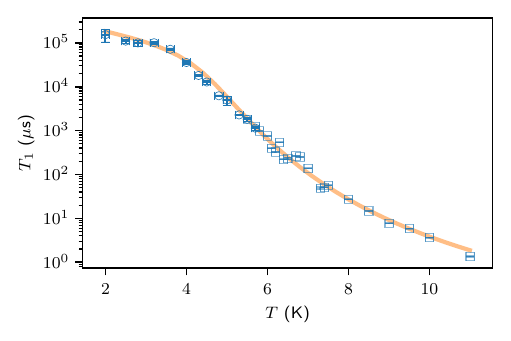}
		\caption{Measured spin-lattice relaxation time $T_1$ of \ce{Tb^3+} in \ce{CaWO4} as a function of temperature (blue dots and squares), extracted by inversion recovery. Solid line, \cref{eq:T1_of_T} with coefficients from Refs.~\cite{breen_spin_lattice_1968,antipin_paramagnetic_1968}. Measurements (dots and squares) were performed in two independent laboratories.}
		\label{fig:T1}
	\end{figure}
	
	The spin-lattice relaxation time $T_1$ sets the ultimate physical limit on coherence through $T_2 \leq 2T_1$. The temperature dependence of $T_1$ for \ce{Tb^3+} in \ce{CaWO4} has previously been obtained from CW linewidth measurements~\cite{breen_spin_lattice_1968,antipin_paramagnetic_1968}:
	\begin{equation}
		\label{eq:T1_of_T}
		\frac{1}{T_1} = \alpha_D T + \alpha_R T^7 + \alpha_O\exp\!\left(-\frac{\Delta_O}{T}\right),
	\end{equation}
	with a direct-process coefficient $\alpha_D = \SI{2.67}{s^{-1}K^{-1}}$, a two-phonon Raman coefficient $\alpha_R = \SI{7.7e-4}{s^{-1}K^{-7}}$, and an Orbach coefficient $\alpha_O = \SI{6.67e8}{s^{-1}}$ with a gap $\Delta_O = \SI{78.8}{K}$ to the first excited crystal-field level~\cite{breen_spin_lattice_1968}. Our measurements, performed by inversion recovery on echo-detected EPR, provide the first direct pulsed determination of $T_1$ over the full temperature range and are in quantitative agreement with \cref{eq:T1_of_T} between 2 and \SI{10}{K} (\cref{fig:T1}). An independent measurement of $T_1$ as a function of field orientation at \SI{6.4}{K} (Appendix \ref{app:T1_orientation}) finds it constant within the \SI{0.05}{K} temperature stability of the cryostat, as expected at fixed spin transition frequency \cite{AbragamBleaney1970}. We therefore treat $T_1$ as orientation-independent in the decoherence model below.
	
	\subsection{Hahn-echo coherence: failure of the spin-bath model}
	\label{sec:T2_failure}
	
	\begin{figure*}[t]
		\centering
		\includegraphics[width=0.95\textwidth]{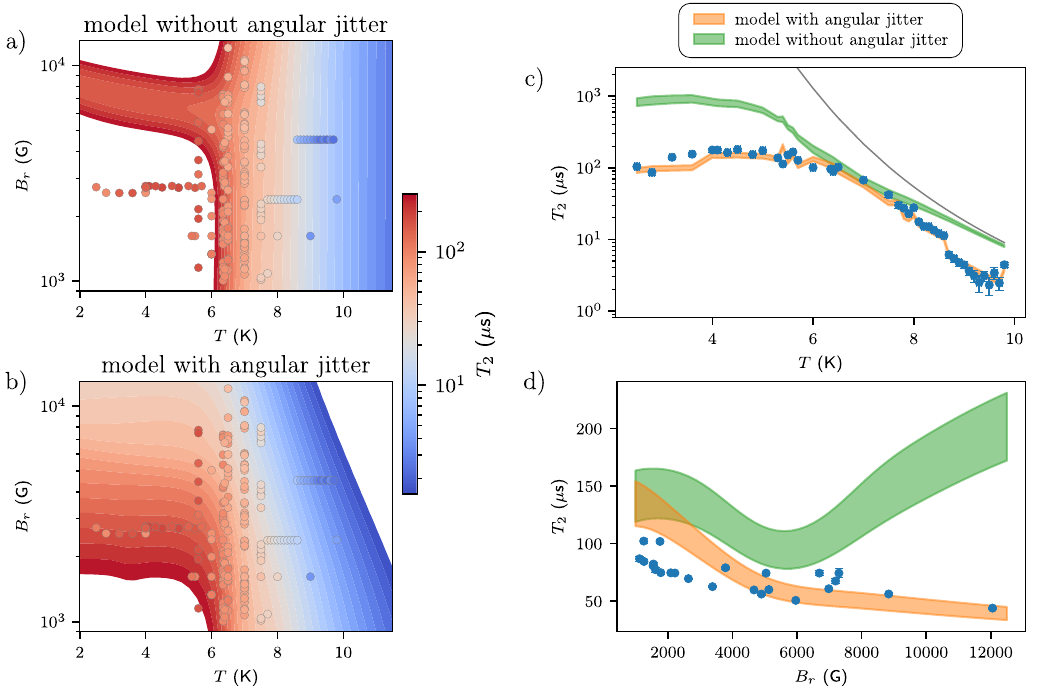}
		\caption{Hahn-echo coherence time $T_2$ of \ce{Tb^3+} in \ce{CaWO4}. \textbf{a,b)} Colored dots, measured $T_2$ as a function of temperature $T$ and resonant field $B_r$. Colored contours, model \textbf{a)} without and \textbf{b)} with angular fluctuations. For visibility, contours outside the colorbar range [$\SI{1.5}{\micro s}, \SI{270}{\micro s}$] are not shown. \textbf{c)} Blue dots, maximum measured $T_2$ at each temperature. Solid gray line, physical limit $2T_1$ from \cref{eq:T1_of_T}. \textbf{d)} Blue dots, measured $T_2$ as a function of resonant field at $T = \SI{6.5}{K}$. In \textbf{c,d)}, green shades, model without angular fluctuations; orange shades, model with angular fluctuations. Shade widths reflect combined uncertainties in temperature (\SI{0.05}{K}) and impurity concentrations (Appendix \ref{app:quantification}). 
        }
		\label{fig:Gamma_m}
	\end{figure*}
	
	We probe decoherence using the Hahn-echo sequence $(\pi/2 - \tau - \pi - \tau - \text{echo})$ as a function of temperature and resonant magnetic field. The coherence time $T_2$ is extracted by fitting the echo decay to a stretched exponential $\exp[-(2\tau/T_2)^\beta]$. To avoid distortions from electron spin echo envelope modulation (ESEEM) from \ce{^183W} nuclear spins, the analysis is restricted to resonant fields $B_r > \SI{1000}{G}$, where the nuclear Zeeman energy sufficiently suppresses the modulation depth (Appendix \ref{app:eseem}). Data were collected for all four \ce{Tb^3+} hyperfine lines and multiple orientations.
	
	Two experimental observations emerge from \cref{fig:Gamma_m}. At high temperature ($T \gtrsim \SI{7}{K}$), $T_2$ approaches $2T_1$ but saturates systematically below it, indicating a pure-dephasing contribution that scales with $T_1$ but that also depends on the resonant field. At low temperature ($T \lesssim \SI{7}{K}$), $T_2$ saturates at 100 to \SI{200}{\micro\second} and \emph{decreases} monotonically with $B_r$ (\cref{fig:Gamma_m}d), a behavior not captured by any spin-bath mechanism.
	
	We first test whether these features can be reproduced by conventional decoherence mechanisms. A parameter-free model is built from spin-lattice longitudinal relaxation (\cref{sec:T1}), instantaneous diffusion, and spectral diffusion from all independently quantified impurity species present in the crystal. The complete set of mechanisms is summarized schematically in Appendix \ref{app:schematic}, and selected numerical values are collected in \cref{tab:Gamma_estimates}. Detailed expressions and all estimated $T_2$ contributions appear in Appendix \ref{app:model}.

    \begin{table*}[ht]
		\centering
		\renewcommand{\arraystretch}{1.3}
		\begin{tabular}{c c c c c c c c c c}
			\hline\hline
			$T$ (K) & $B_r$ (G) & $2T_1$ (ms) & $T_{2}^\mathrm{ID}$ (ms) & $T_{2}^\mathrm{Tb}$ (ms) & $T_{2}^\mathrm{Mo}$ (ms) & $T_{2}^\mathrm{Er}$ (ms) & $T_{2}^\mathrm{W}$ (ms) & $T_{2}^\mathrm{th}$ (ms) & $T_{2}^\mathrm{exp}$ (\si{\micro\second}) \\
			\hline
			2.5  & 2726  & $280 \pm 6$   & $14 \pm 2$  & $3.7 \pm 0.2$   & $1.21 \pm 0.05$       & $2.4 \pm 0.3$ & 16 & $0.77 \pm 0.05$  & $104 \pm 7$  \\
			4    & 2725  & $79 \pm 4$    & $ 14\pm 2$  & $2.7 \pm 0.2$       & $0.96 \pm 0.04$       & $61 \pm 12$ & 16 & $0.76 \pm 0.04$  & $180 \pm 10$ \\
			6.5  & 1263  & $0.50 \pm 0.04$ & $ 2.9\pm 0.3 $ & $0.49 \pm 0.04$      & $0.24 \pm 0.02$  & --  & 12 & $0.14 \pm 0.01$ & $84 \pm 2$  \\
			6.5  & 12054 & $0.50 \pm 0.04$ & $277 \pm 31$   & $0.46 \pm 0.03$ & $1.8 \pm 0.2 $ & -- & 29 & $0.20 \pm 0.01$ & $43.9 \pm 0.6$ \\
			\hline\hline
		\end{tabular}
		\caption{Estimated dominant contributions to $T_2$ for line L4 for the conventional decoherence mechanisms at four representative $(T, B_r)$ points. Contributions above \SI{1}{s} are not shown. $T_2^\mathrm{ID}$: instantaneous diffusion; $T_2^\mathrm{Tb}$: off-resonant terbium spectral diffusion; $T_2^\mathrm{Mo,Er,W}$: spectral diffusion from Mo, Er, and \ce{^183W} respectively; $T_2^\mathrm{th}$: combined model; $T_2^\mathrm{exp}$: measured value. The order-of-magnitude gap between $T_2^\mathrm{th}$ and $T_2^\mathrm{exp}$ at low temperature and the opposite behavior with resonant field are the signatures of the additional decoherence mechanism identified in \cref{sec:jitter}.}
		\label{tab:Gamma_estimates}
	\end{table*}
	
	Instantaneous diffusion arises from the microwave-driven flips of \ce{Tb^3+} spins that share the same nuclear projection $m_I$ as the probed line. Even though the dipolar coupling between terbium ions is independent of orientation, the instantaneous diffusion decreases when $\theta \to \pi/2$ as the inhomogeneous linewidth increases and so the concentration of the probed terbium decreases. In the probed parameter region, its  $T_2$-limit is \SI{2.2 \pm 0.3}{ms} at minimum, \textit{i.e.} at $B_r \geq \SI{1000}{G}$. It is one order of magnitude above the measured values. We verified this experimentally by measuring the echo decay as a function of the $\pi$-pulse flip angle at $B_r = \SI{1620}{G}$ and $T = \SI{6.7}{K}$: no significant change in $T_2$ is observed. 
	
	Direct flip-flops between resonant \ce{Tb^3+} spins require a different analysis than for Kramers ions. With $g_\perp = 0$, the transverse magnetic moment vanishes; the only flip-flop coupling comes from the tunnel-induced off-diagonal matrix element of the \emph{longitudinal} magnetic moment in the energy eigenbasis~\cite{Car2019}. At resonance at X-band, for a fixed cavity frequency $\nu_c$, the Zeeman energy $h \nu_z =h \sqrt{\nu_c^2-\Delta^2}$ is constant and independent of orientation $\theta$ and thus is the flip-flop coupling, a different qualitative behavior than in the Kramers case. For our small native concentration $C_\mathrm{Tb} = \SI{15}{ppb}$, numerical evaluation (Appendix \ref{app:flipflop}) yields a flip-flop lifetime above \SI{e2}{s}, six orders of magnitude longer than the measured $T_2$. This contribution is henceforth ignored.
	
	Spectral diffusion originates from state changes of neighboring spins, which act as a fluctuating dipolar field on the probed spin. Three baths are considered: the off-resonant \ce{Tb^3+} hyperfine lines, the identified Kramers impurities (\ce{Mo^5+}, \ce{Dy^3+}, \ce{Er^3+}, \ce{Ce^3+}, \ce{Nd^3+}), and the \ce{^183W} nuclear spins. State changes proceed mostly through spin-lattice relaxation (indirect-$T_1$ process); indeed, indirect flip-flops within each sub-ensemble are negligible at the measured concentrations. The contribution scales with the spectral diffusion linewidth $\Gamma_\mathrm{SD}$ and with the state-change rate $R = 1/T_1 + R_\mathrm{ff} \simeq 1/T_1$, with three regimes depending on the ratio $R\tau$ (Appendix \ref{app:SD}). 
    The $T_2$ limits set by this mechanism are collected in \cref{tab:Gamma_estimates}. For \ce{^183W}, the phenomenological scaling of Ref.~\cite{Kanai2022}, benchmarked against the cluster-correlation expansion calculation of Ref.~\cite{LeDantec2021} for \ce{Er^3+} in the same host, yields $T_2^\mathrm{W} \gtrsim \SI{12}{ms}$ in our temperature and field range. We also verified that two further effects, the terbium nuclear spin-flip and residual magnet field noise, remain above the \SI{70}{ms} range and are therefore irrelevant (Appendix \ref{app:nuclear}).
	
	The sum of these mechanisms, shown as contours in \cref{fig:Gamma_m}a and as green shades in \cref{fig:Gamma_m}c,d, defines a \emph{parameter-free} reference model with no adjustable quantities. It fails in two independent ways. First, it overestimates the measured $T_2$ at $T = \SI{2.5}{K}$ by almost an order of magnitude. Second, it predicts $T_2$ to be essentially independent of or slightly decreasing and then increasing with $B_r$ in our range of temperature and field, whereas the data tends to show mostly a monotonic decrease. No combination of concentration uncertainties (one standard deviation each), no plausible revision of $T_1$, and no residual unmodeled spin-bath contribution can reproduce a monotonic decrease of $T_2$ with $B_r$ at fixed temperature. A qualitatively new mechanism, with an opposite field dependence, is required.
	
	\subsection{Angular jitter as the dominant decoherence channel}
	\label{sec:jitter}
	
	\begin{figure*}[t]
		\centering
\includegraphics[width=0.95\textwidth]{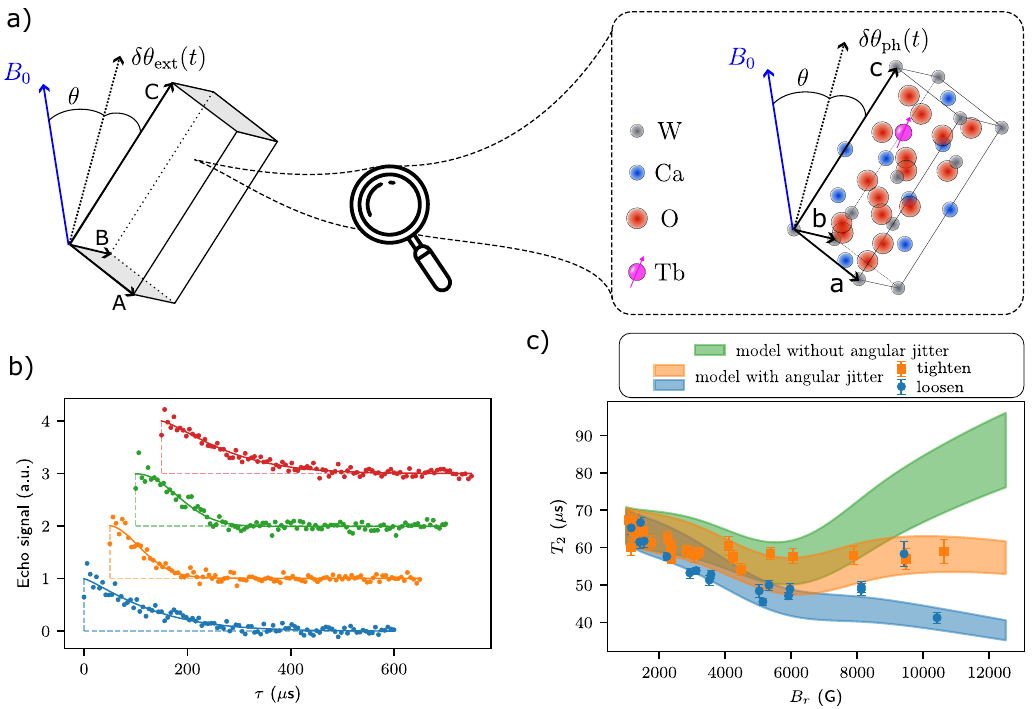}
		\caption{Angular jitter as the dominant decoherence channel. \textbf{a)} Schematic of the two angular-jitter pathways: a global rotation $\delta\theta_\mathrm{ext}$ of the crystal relative to the fixed laboratory field, driven by mechanical vibrations; and a local phonon-driven modulation $\delta\theta_\mathrm{ph}$ of the crystal c-axis at the unit-cell level. \textbf{b)} Hahn-echo decay at $T = \SI{4}{K}$ and $B_r = \SI{2565}{G}$ in four controlled pump configurations. Blue, reference state ($T_2 = \SI{142}{\micro\second}$). Orange, auxiliary pump switched on ($T_2 = \SI{83}{\micro\second}$). Green, same with damping weights on the cryostat tubing ($T_2 = \SI{108}{\micro\second}$). Red, pump switched off, reference restored ($T_2 = \SI{142}{\micro\second}$). Curves are offset vertically for visibility. The signal-to-noise ratio of the echo decay is unaffected throughout, ruling out instrumental artifacts. \textbf{c)} Hahn-echo $T_2$ at $T = \SI{7}{K}$ as a function of $B_r$ for two experimental configurations: setup tightened in orange or setup loosened in blue. Dots, measured $T_2$ in the two configurations. Green shade, model without angular jitter. Blue and orange shades, model with angular jitter, with two different $\delta\theta_\mathrm{ext}$ corresponding to the two different mechanical configurations.}
		\label{fig:jitter}
	\end{figure*}
	
	The missing mechanism is angular jitter: time-varying fluctuations of the angle $\theta$ between $\bm{B}$ and the crystal c-axis. Since $\partial\nu/\partial\theta \neq 0$ for any spin with $g$-factor anisotropy (\cref{eq:sensitivity}), such fluctuations translate directly into transition-frequency fluctuations and hence into pure dephasing. Two physically distinct sources contribute: global mechanical motion of the sample relative to the fixed laboratory field, and phonon-driven local displacements of the crystal axes (\cref{fig:jitter}a).
	
	\subsubsection{Controlled experimental evidence}
	\label{sec:pump_evidence}
	
	Direct and causal evidence for a mechanical origin is provided by two controlled experiments (\cref{fig:jitter}b and c). In the first experiment, at fixed temperature ($T = \SI{4}{K}$) and fixed resonant field ($B_r = \SI{2565}{G}$), switching on an additional pump in the vicinity of the cryostat reduces $T_2$ from \SI{142}{\micro\second} to \SI{83}{\micro\second}. Switching it off restores the initial value. Partial vibration isolation through damping weights on the cryostat tubing recovers $T_2 = \SI{108}{\micro\second}$. The echo-decay signal-to-noise ratio is unaffected throughout (roughly estimated at 18, 19, 17 and 17 respectively), excluding instrumental artifacts from detection, pulse calibration, or thermal drifts. The experiment couples a macroscopic mechanical perturbation to the measured coherence at the tens of percent level, with no intermediate chain of inference.

    In the second experiment, at fixed temperature ($T = \SI{7}{K}$), the measured the Hahn-echo $T_2$ as a function of resonant field $B_r$ in two experimental realizations, one where the setup is maximally tightened, therefore mitigating the transmission of external mechanical vibrations towards the sample and in the second, the setup is loosened, thus increasing the transmission of mechanical noise on the sample. We observe that in the loosened configuration compared to the tightened one, the $T_2$ decreases more as a function of resonant field $B_r$, indicating a stronger angular jitter at the sample level. The model with angular jitter (blue and orange shades in \cref{fig:jitter}.c) indicates an angular noise about twice as large between the two configurations.

    Both controlled experiments establish that the missing decoherence channel is coupled
  to mechanical vibrations. The two experiments deserve one more comment. Rotary and scroll pumps radiate most of their power at their rotation frequency and its low harmonics, well below the
$\sim\SI{2.5}{kHz}$ band probed by the Hahn filter (see \cref{sec:global_jitter}). What reaches the sample is not that
line spectrum but the broadband response of the mechanical assembly to it: the tubing,
the insert and the cryostat structure convert low-frequency forcing into acoustic
excitation of their own resonances, which extend into the kilohertz range in cryogenic
setups~\cite{Kalra2016}. The observation that damping weights placed on the tubing
recover a large fraction of the lost coherence supports this transmission picture rather
than a direct spectral overlap.
	\subsubsection{Hahn-echo filter and the angular noise spectrum}
	\label{sec:global_jitter}
	
	The Hahn-echo sequence acts as a band-pass filter on any frequency-modulating noise. For a stationary angular noise process $\theta(t)$ with one-sided spectral density $S_\theta(\omega)$, assuming a Gaussian noise statistics, the echo envelope coherence function is $C(2\tau) = e^{-\chi(2\tau)}$ with the attenuation function
	\begin{align}
		\label{eq:filter_integral}
       \chi(2\tau) &= \frac{1}{2\pi}\int_0^\infty  S(\omega)\,F_H(\omega, 2\tau)\,d\omega \notag \\
       & =
       \,\frac{1}{2\pi}\int_0^\infty \left(2\pi\,\pdv{\nu}{\theta}\right)^{\!2} S_\theta(\omega)\,F_H(\omega, 2\tau)\,d\omega,
	\end{align}
	where $S(\omega)$ the noise power spectral density in $\si{(rad/s)^2 / (Hz)}$, and $F_H(\omega, 2\tau) = \sin^4(\omega2\tau/4)/(\omega/4)^2$ the Hahn filter function. $F_H(\omega, 2\tau)$ vanishes at $\omega = 0$ and peaks near $\omega \sim \pi/\tau$, so the Hahn echo refocuses static and quasi-static frequency offsets and probes mostly the noise content within a band of width $\sim 1/\tau$ around this peak.

Determining the full noise density $S_{\theta}(\omega)$ is beyond the scope of this manuscript. However, assuming the simplification of a $1/f$ type of noise $S_\theta(\omega) = \frac{2\pi \delta\theta^2}{\omega}$, which is often encountered in solid-state system \cite{Paladino2014}, the Hahn-echo attenuation function becomes \cite{Ithier2005} $\chi(2\tau) = (\Gamma_\mathrm{\theta, eff} \tau)^2 = (2\sqrt{\ln(2)} \pdv{(2\pi\nu)}{\theta}\, \delta\theta \tau)^2$ resulting in a Gaussian decay with an effective dephasing time
        \begin{equation}
        \label{eq:eff_theta}
            T_{2, eff} = \frac{2}{\Gamma_\mathrm{\theta, eff}} = \frac{1}{2\pi\sqrt{\ln(2)} \pdv{\nu}{\theta}\, \delta\theta}
        \end{equation}
        
	From \cref{eq:sensitivity}, $T_{2, eff}^{-1}$ grows linearly with $B_r$ for $B_r \gtrsim \SI{1000}{G}$, accounting directly for the observed monotonic decrease of $T_2$ with resonant field (\cref{fig:Gamma_m}d).  The measured stretched-exponential exponent $\beta = 1.8 \pm 0.3$ extracted in the angular-jitter-dominated regime (\cref{app:beta}) at low temperature justifies this 1/f approximation for the noise spectral density. With the Hahn echo filter $F_H(\omega, 2\tau)$ which mostly probes the noise at the rate $\omega_\tau \sim \frac{\pi}{\tau}$, the \ce{Tb^3+} spins are sensitive to this $1/f$ type of noise in a frequency window typically above $f_{\tau=T_2} = \frac{\omega_{\tau=T_2}}{2\pi} \sim \frac{1}{2T_2} \sim \SI{2.5}{kHz}$ for $T_2 \sim \SI{200}{\micro s}$. 
    
    This $1/f$ type of noise is only an approximation. In practice, there are cutoff frequencies $\omega_\mathrm{low}$ (given by how long we measure) and $\omega_\mathrm{high}$ (given by the measurement bandwidth) in between which the noise is of type $1/f$. These cutoff frequencies could also come from the noise process itself. The resulting root-mean-square noise amplitude is given by $\delta\theta_\mathrm{rms} = \sqrt{\frac{1}{2\pi} \int_{\omega_\mathrm{low}}^{\omega_\mathrm{high}} S_\theta(\omega) d\omega } = \delta\theta \sqrt{\ln(\frac{\omega_\mathrm{high}}{\omega_\mathrm{low}})}$.  
    Taking one over the total time of the measurement for $\omega_\mathrm{low}$ and for $\omega_\mathrm{high}$ the measurement bandwidth, which is about GHz frequency, results in $\delta\theta_\mathrm{rms} \sim 5 \delta\theta$.
     
    	\subsubsection{Global and local phonon-driven angular jitter}
	\label{sec:local_jitter}

    Inside the probed angular noise $S_\theta(\omega)$ with the Hahn-echo sequence, we distinguish two sources (\cref{fig:jitter}.a). First, a temperature-independent contribution $ \delta\theta_\mathrm{ext}$ assumed to come from external mechanical vibrations that make the sample globally change orientation. With this assumption, we neglect any changes of transmission of mechanical vibration depending on the temperature of each element of the setup from the source of mechanical vibration towards the sample. Including this source in the spin bath model allows to explain the saturation of $T_2$ around \SIrange{100}{200}{\micro\second} at low-temperature. 
    However, it does not reproduce the behavior in the high-temperature limit, where a second temperature-dependent source of angular jitter is necessary.

Indeed, at high temperature ($T \gtrsim \SI{7}{K}$), the observed $T_2(T)$ approaches but does not reach $2T_1(T)$ (see \cref{fig:Gamma_m}.c and Appendix \ref{app:jitter_phonons}). This effect is not explained by our parameter-free spin bath model nor by the global angular jitter contribution. Not only $T_2(T)$ does not reach $2T_1(T)$ but the leftover dephasing $1/T_2- 1/(2T_1)$ scales with $1/T_1$. In this temperature region, $T_1$ is mostly dominated by the Orbach process. An Orbach process with the same gap $\Delta_O = \SI{78.8}{K}$ can fit the extra dephasing in this temperature region (see Appendix \ref{app:jitter_phonons}). This signals that a spin-phonon pure-dephasing contribution has to be added to the usual spin-phonon longitudinal contribution $T_1(T)$ \cite{Bar-Gill2013,Mondal2023}. This pure dephasing contribution has been computed for NV centers and can result in a $T_2$ that scales with $T_1$ \cite{Bar-Gill2013,Mondal2023}. Moreover, the scaling of the extra dephasing with $1/T_1$ scales linearly with the resonant field or equivalently with the angular sensitivity $\pdv{\nu}{\theta}$ in this high-temperature limit and in our probed region of resonant field.  

We empirically assume that the Hahn-echo decay due to this temperature-dependent source takes the same form as the global contribution due to external mechanical noise. We then write the effective decay rate
	\begin{align}
    \label{eq:gamma_theta_phonon}
		\Gamma_\mathrm{ph, eff} &= 2 \sqrt{\ln(2)} \,\left|\pdv{(2\pi\nu)}{\theta}\right| \delta\theta_{ph} \notag \\
        &= 2 \sqrt{\ln(2)} \,\left|\pdv{(2\pi\nu)}{\theta}\right|\frac{\alpha_\mathrm{ph}}{T_1(T)},
	\end{align}
    where we have adopted the same functional form as \cref{eq:eff_theta} but with a temperature-dependent effective amplitude $\delta\theta_{ph}(T) = \frac{\alpha_\mathrm{ph}}{T_1(T)}$, which encodes the empirical observation that this decoherence channel increases with the spin-lattice rate $1/T_1(T)$, while also increasing with $B_r$ through $\pdv{(2\pi\nu)}{\theta}$. 
 This phenomenological form above suffices to match the high-temperature dataset with the single coefficient $\alpha_\mathrm{ph}$.

    With the assumption that this effect comes from spin-phonon pure dephasing, we interpret it as a phonon-driven local modulation of the crystal c-axis at the unit-cell level. 
    Thus, \cref{eq:gamma_theta_phonon} captures this local angular jitter contribution with the correct temperature and resonant field scalings, while remaining model-independent in its detailed spin-phonon couplings and phonon spectral content.
    A microscopic derivation from first principles \cite{Mondal2023} is beyond the scope of this work but should be performed to validate our interpretation and rule out other possible interpretation. 
    
    We treat the two angular jitter noise contributions as independent so that we add the corresponding attenuation function $\chi(2\tau)$, i.e. we add the $\Gamma_\mathrm{eff}^2$, so we have a corresponding total angular jitter amplitude $\delta\theta_\mathrm{tot} = \sqrt{\delta\theta_\mathrm{ext}^2+\delta\theta_\mathrm{ph}^2}$. In the low temperature region $T<\SI{7}{K}$, the global angular jitter contribution is the dominant one $\delta\theta_\mathrm{ext}\gg \delta\theta_\mathrm{ph}$ while reciprocally, in the high temperature region $T>\SI{7}{K}$, the temperature-dependent contribution assumed to be locally phonon-induced angular jitter is the dominant one $\delta\theta_\mathrm{ph}\gg \delta\theta_\mathrm{ext}$.

	\subsubsection{Complete model and global fit}
	\label{sec:complete_model}
	
	The complete model includes all mechanisms of \cref{sec:T2_failure} together with \cref{eq:eff_theta,eq:gamma_theta_phonon}. Only two scalar parameters are free, $\delta\theta_\mathrm{ext}$ and $\alpha_\mathrm{ph}$; all other quantities (concentrations, spin-flip rates, $T_1$) are independently measured or taken from the literature. A global fit to the full temperature and resonant-field dataset yields
	\begin{equation}
		\delta\theta_\mathrm{ext} = \SI{1.8}{\micro\degree}, \qquad \alpha_\mathrm{ph} = \SI{0.28}{\milli\degree\cdot\micro\second}.
	\end{equation}
	As shown by the contours in \cref{fig:Gamma_m}b and the orange shades in \cref{fig:Gamma_m}c,d, the model reproduces the measured $T_2$ across the full ranges of temperature (\SIrange{2}{10}{K}) and resonant field ($10^3$ to $\SI{e4}{G}$), across multiple runs, setups, and laboratories. The global angular jitter explains the low-temperature plateau at $T_2 \sim \SI{150}{\micro\second}$ below \SI{6}{K}; the local phonon-driven jitter captures the $T_2 < 2T_1$ behavior at higher temperature. The linear increase of dephasing with $B_r$, which was the spin-bath model's fatal flaw, emerges naturally from \cref{eq:sensitivity,eq:eff_theta,eq:gamma_theta_phonon}.

	From our model, we can decompose the transverse relaxation rate into its individual contributions (see Appendix \ref{app:Gamma_contributions}). At $T = \SI{6.5}{K}$, angular jitter dominates at $B_r \gtrsim \SI{3000}{G}$ and is the only contribution growing with field; $2T_1$ and \ce{Mo^5+} spectral diffusion are subleading and field-decreasing. The shades in \cref{fig:Gamma_m}c,d include a \SI{10}{\percent} relative uncertainty on $\delta\theta_\mathrm{ext}$ and $\alpha_\mathrm{ph}$ in addition to the uncertainties on $T$ and concentrations.
	
	The lower $T_2$ values observed below \SI{4}{K}, where an auxiliary helium pump must be activated to reach these temperatures, are reproduced by increasing $\delta\theta_\mathrm{ext}$ to \SI{3.3}{\micro\degree}, consistent with the controlled pump experiment of \cref{sec:pump_evidence}. The estimated values of $\delta\theta_\mathrm{ext}$ from the $B_r$ dependence at fixed $T$ can vary from run to run and from setup to setup (\cref{fig:jitter}b-c and Appendix \ref{app:setup_jitters}), with an upper bound of \SI{2.3}{\micro\degree} under normal operating conditions.
	
	\subsection{Ruling out alternative explanations}
	\label{sec:alternatives}
	
	Before moving to the sensing discussion, we explicitly address possible alternatives to the angular-jitter interpretation. (i) \emph{Calibration of impurity concentrations.} Increasing each impurity concentration within its one-standard-deviation envelope (Appendix \ref{app:quantification}) shifts the green shade in \cref{fig:Gamma_m}c,d by less than a factor two and cannot produce the observed \emph{decrease} of $T_2$ with $B_r$. (ii) \emph{Instrumental pulse phase noise.} At long interpulse delays, instrumental phase fluctuations of the $\pi$-pulse may lead to an apparent reduction of $T_2$~\cite{Tyryshkin_2006}. This artifact is suppressed by magnitude detection of the echo, which we use throughout. Since we observe no change in the extracted $T_2$, we exclude this mechanism. (iii) \emph{Instrumental magnetic noise.} An isotropic field-noise contribution is negligible as it overestimates by more than two orders of magnitude the observed $T_2$. More decisively, it would enter through
$\partial\nu/\partial B \propto \cos\theta$, hence \emph{decrease} with $B_r$, which is the opposite of the measured trend. (iv) \emph{Tb nuclear spin flips.} Assuming slow nuclear dynamics ($T_{1,n} \gg T_1$), this contribution is negligible; even if $T_{1,n}$ were short, it would not reproduce the linear field dependence of $1/T_2$. 
    (v) \emph{Unmodeled fluctuating spin bath.} It is still possible that some spin species in the bath has been overlooked. However, the required concentration and necessary $T_2$-induced behavior by spectral diffusion at low $T$ and as a function of resonant field, make this hypothesis improbable.
    Finally, the two controlled experiments (\cref{sec:pump_evidence}) further rule out any mechanism not coupled to mechanical vibrations.
	
	\section{Spin-coherence-based angular probe}
\label{sec:sensing}

The same property that makes \ce{Tb^3+} vulnerable to mechanical noise, the diverging angular sensitivity $\partial\nu/\partial\theta$ as $\theta \to \pi/2$, also turns it into a transducer of orientational fluctuations. Approaching the ab-plane drives the angular-jitter dephasing rate upward without bound while spin-bath contributions are mostly suppressed or remain constant. Within the rigorous $g_\perp = 0$ result (\cref{sec:sample}), an orientation and resonant field can therefore always be chosen at which $T_2$ is limited by angular noise alone, however small the angular jitter amplitude $\delta\theta_{tot}$.

The natural figure of merit is the angular noise spectral density that the probe extracts from its environment. With our assumption of a 1/f type of noise, we directly extract a value of 
\begin{equation}
\label{eq:S_theta_sensing}
\sqrt{S_\theta} \approx \SI{36}{\nano\degree/\sqrt{Hz}}.
\end{equation}
for the global external mechanical contribution at the lowest estimated frequency $1/(2T_2) \sim \SI{2.5}{kHz}$. This is the present estimated sensitivity of the device: the angular noise of the spectrometer/cryostat setup itself, sampled through the Hahn-echo filter. The relative displacement implied by $\delta\theta_\mathrm{ext} \approx \SI{1.8}{\micro\degree}$ at the sample scale $L\sim \SI{2}{mm}$, $L\,\delta\theta \sim \SI{0.06}{nm}$, is well below the floor of non-isolated lab environments and consistent with the residual jitter of a well-mounted cryostat coupled to the acoustic background, scroll pumps, and helium circulation. 

The standard-model contributions to $T_2$ at $T = \SI{2.5}{K}$ and $B_r = \SI{2726}{G}$ give a coherence floor $T_2^\mathrm{floor} \approx \SI{0.8}{ms}$ (\cref{tab:Gamma_estimates}), almost an order of magnitude above the currently observed $T_2$. With dedicated mechanical isolation reducing the environmental contribution to that coherence floor, the achievable resolution would scale to $\sqrt{S_\theta^\mathrm{floor}} \sim \SI{5}{\nano\degree/\sqrt{Hz}}$, before any instrumental noise floor of the spectrometer becomes relevant.

Two practical limits qualify these numbers. The first is geometric: approaching $\theta = \pi/2$ requires fine angular control of the sample, while the inhomogeneous linewidth grows together with the angular sensitivity (\cref{fig:EFS}b), reducing the fraction of spins within the cavity excitation bandwidth. The second is the clock-transition trade-off: as $\nu_c \to \Delta$, both the magnetic and the angular sensitivities are suppressed by the common factor $\sqrt{\nu_c^2 - \Delta^2}/\nu_c$ (Appendix \ref{app:clock_transition}). Coherence is then protected against angular jitter and spin bath magnetic noise at the cost of reduced sensing gain, defining a parameter space in which angular detection and magnetic-noise immunity may be jointly optimized.

It is informative to compare with other mechanical sensors. Among those sensors, different physical quantities like acceleration, velocity, angular velocity, displacement or angle/orientation are detected. Two distinct physical quantities cannot be compared in absolute units without some geometric assumptions that we deliberately do not make. We note nonetheless a related rare-earth sensor based on optical coherence in \ce{Tm}:YAG \cite{louchet_chauvet_2022} that has demonstrated a \SI{e-16}{m/\sqrt{Hz}} displacement sensitivity at \SI{1}{MHz} with \SI{0.7}{\micro W} of optical power. For the detection of angle/orientation, to our knowledge, the literature is relatively scarce. In the low frequency band ($\lesssim \SI{100}{Hz}$), several hardwares with different sensitivity have been reported : electromagnetically induced transparency in Rb vapor with  $\SI{7}{\milli\degree/\sqrt{Hz}}$ sensitivity \cite{GonzalezMaldonado2024}, heterodyne interferometer of laser beam with $\SI{6}{\nano\degree/\sqrt{Hz}}$ \cite{Hahn2010} or tiltmeter  with $\SI{40}{\pico\degree/\sqrt{Hz}}$ sensitivity \cite{Allocca2021}. However, these values are hard to compare to our anisotropic non-Kramers spin detection scheme as the Hahn echo filtering bandwidth makes our detection not sensitive to noise in this low frequency band. Very few angular sensitivity values have been reported in a higher frequency band, Ref. \cite{PAOLINO2007} reports an integrated noise of \SI{0.68}{nrad}
for frequency up to a few kHz with single laser beam interferometric detection while Ref. \cite{Qiu2021} reports $\SI{13}{\milli\degree/\sqrt{Hz}}$ sensitivity in an estimated frequency band around \SI{100}{kHz}. Our sensor thus fills a gap and provides a complementary sensing tool for mechanical noise above $\sim \SI{2.5}{kHz}$ in cryogenic setups where mechanical noise extending up to the high kHz frequency band can be expected \cite{Kalra2016}.

This angular jitter mechanism is not specific to \ce{Tb^3+} in \ce{CaWO4}. It applies to any spin transition with $g$-factor anisotropy and is particularly sharp for non-Kramers ions, where the symmetry-imposed $g_\perp = 0$ already produces a divergent $\partial\nu/\partial\theta$ at moderate orientations. Angular jitter should therefore be considered in any precision coherence experiment involving anisotropic spin systems, especially in setups equipped with cryostat actuators, piezoelectric positioners, or other mechanical elements close to the sample. A terbium-doped crystal mounted on a mechanical resonator~\cite{Kolkowitz2012} or driven by engineered phonon modes~\cite{greffe2026} would provide a coherent interface between angular motion and spin dynamics, with bidirectional spin-phonon transduction as a natural extension.
	
	\section{Conclusion}
	
	We have reported the first pulsed-EPR measurements of the quantum dynamics of \ce{Tb^3+} in \ce{CaWO4}, from 2 to \SI{10}{K} and over one decade of resonant field. The measured longitudinal relaxation agrees quantitatively with the CW-derived model of Refs.~\cite{breen_spin_lattice_1968,antipin_paramagnetic_1968}. The transverse relaxation exhibits two features that are inaccessible to any standard spin-bath model: a low-temperature plateau at $T_2 \sim \SI{150}{\micro\second}$, well below the $2T_1$ limit, and a monotonic decrease of $T_2$ with resonant field. A two-parameter extension incorporating global and local angular jitter of the crystal c-axis accounts for the complete dataset across setups and laboratories. Two controlled experiments with different mechanical configurations establish the mechanical origin of the dominant contribution.
	
	The dual nature of this coupling, decoherence channel and angular probe, reframes mechanical noise as an intrinsic aspect of quantum experiments with anisotropic spins. Tight mechanical control is essential in setups targeting long coherence times with non-Kramers ions; conversely, the same ions resolve angular noise at the few-tens-of-\si{\nano\degree\per\sqrt{Hz}}
level above the kHz band. In future work, dynamical decoupling noise spectroscopy~\cite{Alvarez2011} will enable complete mapping of the angular noise spectrum $S_\theta(\omega)$, revealing noise sources and mitigation pathways.  
    Finally, hybrid quantum systems integrating terbium-doped crystals into mechanical nano- or microstructures~\cite{Kolkowitz2012,greffe2026} could exploit the effect reported here for bidirectional spin-phonon transduction and sensing.
	
	\begin{acknowledgments}
		 Financial support from the IR INFRANALYTICS FR2054 for conducting the research is gratefully
acknowledged. This work was funded, in part, by the French National Research Agency (ANR) under projects HighpointEPR “ANR-24-CE29-6059-01”,  Qmemo "ANR-22-PETQ-0010" A.M.K. is supported by ANR QuantEdu-France (22-CMAS0001) and by France 2030 investment plan, as part of the Initiative d’Excellence d’Aix-Marseille Université - SpinEClock project A*MIDEX (AMX-22-RE-AB-199). We are grateful to A. Savoyant, V. Dolocan and M. Kuzmin for fruitful discussions.
	\end{acknowledgments}
	
	\bibliography{bibliob.bib}
	\clearpage
	\onecolumngrid
	
\appendix
\crefname{section}{Appendix}{Appendices}
\Crefname{section}{Appendix}{Appendices}

	\section{\ce{Tb^3+}:\ce{CaWO4} microscopic structure}
	
	\subsection{Ground states}
\label{app:ground}

\emph{Crystal-field mixing.} The $S_4$ crystal field couples states whose
$J_z$ quantum numbers differ by four. Within the $J = 6$ ground manifold,
it therefore mixes $|J_z = +6\rangle$ with $|J_z = +2\rangle$, and symmetrically for the time-reversed partners. The
two lowest eigenstates are~\cite{forrester_paramagnetic_1962,kirton1965}
\begin{align}
	|\phi_1\rangle &= \alpha\,|+6\rangle + \beta\,|+2\rangle, \\
	|\phi_2\rangle &= \alpha\,|-6\rangle + \beta\,|-2\rangle,
\end{align}
both belonging to the same singlet representation ${}^1\Gamma_2$ of $S_4$.
The coefficients follow directly from the measured $g_\parallel$: for a
state of composition $\alpha|+6\rangle + \beta|+2\rangle$, one has
$g_\parallel = 2 g_J(6\alpha^2 - 2\beta^2) = 3(6\alpha^2 - 2\beta^2)$,
which for $g_\parallel = 17.777$ yields $\alpha^2 = 0.991$ and
$\beta^2 = 0.009$, close to a pure $|J_z = \pm 6\rangle$
pair~\cite{forrester_paramagnetic_1962}.

\emph{Origin of the tunnel splitting.} $|\phi_1\rangle$ and
$|\phi_2\rangle$ are exchanged by time reversal, but \ce{Tb^3+} has an
even number of electrons: Kramers' theorem does not protect their
degeneracy. Since the two states transform as the same one-dimensional
representation, the crystal field possesses a non-zero matrix element
$\langle\phi_1|\hat{H}_\mathrm{CF}|\phi_2\rangle = h\Delta/2$ connecting
them. Including the Zeeman interaction along the c-axis, the Hamiltonian
restricted to the quasi-doublet reads, in the
$\{|\phi_1\rangle, |\phi_2\rangle\}$ basis,
\begin{equation}
	\hat{H}/h = \frac{1}{2}
	\begin{pmatrix}
		\nu_z & \Delta \\
		\Delta & -\nu_z
	\end{pmatrix},
	\qquad
	h\nu_z = g_\parallel \mu_B B\cos\theta,
	\label{eq:H2x2}
\end{equation}
which is precisely the effective spin-$1/2$ form of \cref{eq:Hamiltonian}
(the hyperfine term simply shifts $\nu_z \to \nu_z + A_\parallel m_I$). At
zero field, the eigenstates are the symmetric and antisymmetric
combinations
\begin{align}
	|+\rangle &= \frac{|\phi_1\rangle + |\phi_2\rangle}{\sqrt{2}}, &
	|-\rangle &= \frac{|\phi_1\rangle - |\phi_2\rangle}{\sqrt{2}},
\end{align}
with energies $\pm h\Delta/2$; the zero-field gap is the tunnel splitting
$\Delta = \SI{8.136}{GHz}$~\cite{forrester_paramagnetic_1962}. In the
effective spin-$1/2$ language, $|\phi_1\rangle$ and $|\phi_2\rangle$ are
the eigenstates of $\hat{S}_z$ and the crystal-field coupling is the
$\Delta\hat{S}_x$ term of \cref{eq:Hamiltonian}.

\emph{Consequences for this work.} Three properties of the quasi-doublet
are central to the main text. First, the near-unity weight of
$|J_z = \pm 6\rangle$ ($\alpha^2 > \SI{99}{\percent}$) produces the giant
longitudinal $g$-factor $g_\parallel \simeq 17.777$, which makes the
transition frequency, and therefore the coherence, exquisitely sensitive
to the field orientation. Second, the small admixture
$\beta \approx 0.097$ enables microwave transitions: the matrix element
coupling $|+\rangle$ and $|-\rangle$ through $B_1 \parallel c$ is
proportional to $\alpha\beta$ and would vanish for a pure
$|J_z = \pm 6\rangle$ pair~\cite{forrester_paramagnetic_1962,kirton1965};
the experimental geometry this imposes is discussed in
Appendix~\ref{app:pulseEPR}. Third, since the quasi-degeneracy is set
entirely by the crystal field rather than by time-reversal symmetry, it
responds to any perturbation of the local environment: phonons, strain,
and changes in the orientation $\theta$ of the applied field. This is the
microscopic origin of the angular-jitter decoherence identified in the
main text.
	
	\subsection{CW spectra and orthorhombic sites}
	\label{app:cw}
	
	\begin{figure}[!ht]
		\centering
		\includegraphics[width=8.6cm]{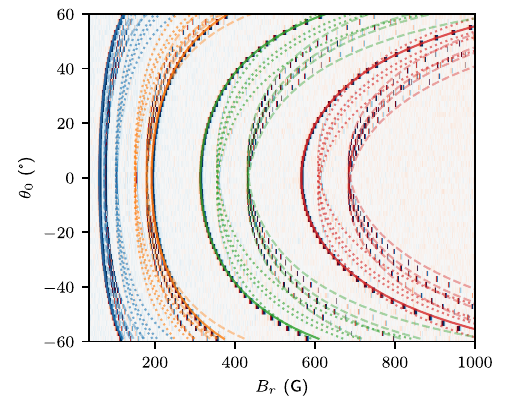}
		\caption{Colorplot, measured CW-EPR spectra in parallel mode. The colorscale is logarithmic to enhance the visibility of the small lines. Computed \ce{Tb^3+} EPR lines (blue L1, orange L2, green L3, red L4) for $S_4$ symmetry (solid lines), orthorhombic site 1 with $\Delta_{o1} = \SI{5.49}{GHz}$ (dashed lines), and orthorhombic site 2 with $\Delta_{o2} = \SI{7.44}{GHz}$ (dotted lines).}
		\label{fig:cw_ortho}
	\end{figure}
	
	Using an X-band continuous-wave EPR spectrometer (Bruker ELEXSYS) with a dual-mode cavity in parallel mode, we measure the CW spectra of \ce{Tb^3+} as a function of the rotation angle $\theta$ (see \cref{fig:EFS,fig:cw_ortho}). The four transitions of the effective spin-$1/2$ are well reproduced by \cref{eq:Bres} and allow the orientation of the sample to be extracted from the observed resonant fields. Two additional sets of four terbium lines are visible at weaker intensity (\cref{fig:cw_ortho}). They are attributed to orthorhombic sites and are well fitted by assuming four possible angle offsets $\theta_{o} = (-6, -1, +1, +6)^\circ$ and modified tunnel splittings $\Delta_{o1} = \SI{5.49}{GHz}$ and $\Delta_{o2} = \SI{7.44}{GHz}$. Their concentrations are estimated at about \SI{3}{\percent} (site 1) and \SI{1}{\percent} (site 2) of the dominant $S_4$ terbium population, so their contribution to the Hahn-echo decay is neglected in the main-text model.
	
	\begin{figure}[!ht]
		\centering
		\includegraphics[width=8.6cm]{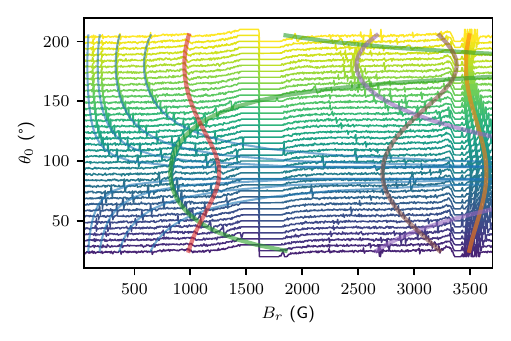}
		\caption{Measured CW-EPR spectra in perpendicular mode at \SI{13}{K} as a function of magnetic field and rotation in the ac-plane. Each line is offset vertically according to its orientation $\theta$. Colored lines: computed resonant fields for \ce{Tb^3+} (blue), \ce{Mo^5+} (orange), \ce{Er^3+} (green), \ce{Dy^3+} (red), \ce{Ce^3+} (purple) and \ce{Nd^3+} (brown). Terbium lines remain visible because of a small sample tilt.}
		\label{fig:spectra_perp}
	\end{figure}
	
	Perpendicular-mode CW spectra (\cref{fig:spectra_perp}) reveal additional Kramers impurities through their angular dependence: \ce{Mo^5+}, \ce{Dy^3+}, \ce{Er^3+}, \ce{Ce^3+} and \ce{Nd^3+}. Their $g$-tensors and concentrations are quantified in Appendix \ref{app:quantification} and summarized in \cref{tab:concentration}.
	
	\subsection{Quantification of Kramers paramagnetic impurities}
	\label{app:quantification}
	
	We calibrate the CW-EPR intensity against an \ce{Er^3+}-doped \ce{CaWO4} reference crystal with previously determined concentration $C_\mathrm{Er}^\mathrm{ref} = \SI{0.70 \pm 0.14}{ppb}$~\cite{LeDantec2021}. Impurity concentrations are extracted from the ratio of double-integrated EPR intensities, normalized by sample mass, measurement temperature, field-modulation amplitude, and transition probability~\cite{quantitative_epr_eaton}. Propagating the reference uncertainty together with additional contributions from mass determination, temperature stability, and fitting error, the erbium concentration in our crystal is
	\begin{equation}
		C_\mathrm{Er} = \SI{4.5 \pm 0.9}{ppb},
	\end{equation}
	which is then used as the internal standard. The remaining Kramers ions are
	\begin{align}
		C_\mathrm{Mo^{5+}} &= \SI{0.30 \pm 0.07}{ppm}, \\
		C_\mathrm{Dy^{3+}} &= \SI{0.07 \pm 0.01}{ppm}, \\
		C_\mathrm{Ce^{3+}} &= \SI{0.7 \pm 0.3}{ppb}, \\
		C_\mathrm{Nd^{3+}} &= \SI{0.5 \pm 0.3}{ppb}.
	\end{align}
	
	\subsection{Quantification of terbium}
	\label{app:quantification_Tb}
	
	Quantification of non-Kramers \ce{Tb^3+} against a Kramers \ce{Er^3+} reference requires dedicated treatment~\cite{hendrich_integer-spin_1989}. Owing to the extreme anisotropy of \ce{Tb^3+} and the associated selection rules, its transitions are absent from the standard perpendicular-mode configuration when the c-axis is aligned with the static field. We introduce a controlled tilt $\theta_t \approx 30^\circ$ between the c-axis and $\bm{B}_0$, determined from the angular dependence of the resonance field, which enables detection and quantification in perpendicular mode. Correcting for both the slope of the energy-field dispersion and the transition matrix element ($|\langle i|g_\parallel\sin\theta_t\,S_z|j\rangle|^2$ for terbium and $|\langle i|g_{\perp,\mathrm{Er}}\cos\theta_t\,S_x - g_{\parallel,\mathrm{Er}}\sin\theta_t\,S_z|j\rangle|^2$ for erbium), the extracted terbium concentration is
	\begin{equation}
		C_\mathrm{Tb} = \SI{15 \pm 4}{ppb}.
	\end{equation}
	
	\begin{table}[!ht]
		\centering
		\begin{tabular}{lccc}
			\hline\hline
			Ion & $g_\parallel$ & $g_\perp$ & Concentration \\
			\hline
			\ce{Tb^3+} & 17.777 & 0 & \SI{15 \pm 4}{ppb} \\
			\ce{Er^3+} & 1.247 & 8.38 & \SI{4.5 \pm 0.9}{ppb} \\
			\ce{Ce^3+} & 2.92 & 1.43 & \SI{0.7 \pm 0.3}{ppb} \\
			\ce{Nd^3+} & 2.03 & 2.52 & \SI{0.5 \pm 0.3}{ppb} \\
			\ce{Dy^3+} & 7.267 & 5.466 & \SI{0.07 \pm 0.01}{ppm} \\
			\ce{Mo^5+} & 1.987 & 1.887 & \SI{0.30 \pm 0.07}{ppm} \\
			\hline\hline
		\end{tabular}
		\caption{$g$-tensors and concentrations for the detected and quantified ions in the studied \ce{CaWO4} crystal. The tensors are axially symmetric, with $g_\parallel$ aligned along the c-axis and $g_\perp$ in the $(a,b)$ plane. For \ce{Tb^3+}, $g_\perp = 0$ is the rigorous consequence of time-reversal symmetry and $S_4$ site symmetry~\cite{griffith1963}.}
		\label{tab:concentration}
	\end{table}
	
	\section{Clock transition}
	\label{app:clock_transition}
	
	\begin{figure}[!ht]
		\centering
		\includegraphics[width=8.6cm]{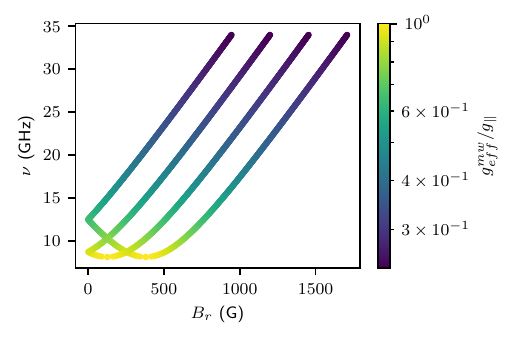}
		\caption{Computed normalized effective microwave $g$-factor $g^\mathrm{mw}_\mathrm{eff}/g_\parallel$ as a function of magnetic field and frequency for $\theta = 0$. The contour at $g^\mathrm{mw}_\mathrm{eff}/g_\parallel \to 1$ marks clock transitions where $\partial\nu/\partial B = 0$.}
		\label{fig:clock}
	\end{figure}
	
	The effective Hamiltonian of \cref{eq:Hamiltonian} features a tunnel splitting $\Delta$ that induces clock transitions at specific points of the $(\nu, B_r)$ plane where the derivative of the spin transition frequency with respect to magnetic field vanishes, $\partial\nu/\partial B \to 0$. Near these points the transition is, to first order, immune to magnetic-field fluctuations. The effective microwave $g$-factor $g^\mathrm{mw}_\mathrm{eff}$, obtained by rescaling $g_\parallel$ with the transition matrix $2|\langle\downarrow|S_\parallel|\uparrow\rangle|$, controls the microwave power required for a given operation (scaling as $(g^\mathrm{mw}_\mathrm{eff})^2$). Near the clock transition, $g^\mathrm{mw}_\mathrm{eff}$ approaches $g_\parallel = 17.777$, almost nine times larger than a typical $g \sim 2$. About \SI{20}{dB} less microwave power is therefore required to drive the same pulse as for a spin with $g \sim 2$ of the same duration.
	
	The suppression factor $\sqrt{\nu_c^2 - \Delta^2}/\nu_c$ shared by \cref{eq:sensitivity,eq:Gamma_dd} vanishes at the clock transition. The angular sensitivity $\partial\nu/\partial\theta$ vanishes together with the magnetic sensitivity: the clock transition protects coherence against magnetic noise and against angular fluctuations. This offers an alternative operating point at which angular jitter is intrinsically suppressed, at the cost of reduced sensing gain.
	
	\section{Pulsed EPR of non-Kramers ions}
	\label{app:pulseEPR}
	\label{app:setup}
	
	\begin{figure}[!ht]
		\centering
		\includegraphics[width=8.6cm]{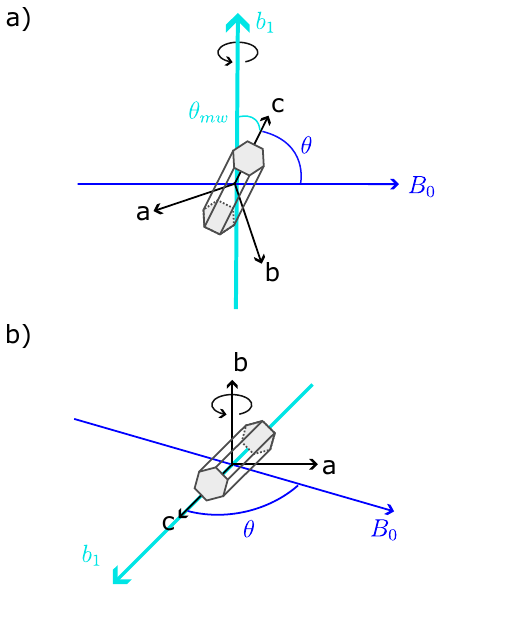}
		\caption{\textbf{a)} Commercial Bruker setup with MD5 cavity. The microwave $b_1$ field is vertical and perpendicular to $B_0$; both are fixed in the laboratory frame. The sample is tilted by an angle $\theta_\mathrm{mw}$ to obtain a non-zero projection of $b_1$ along the c-axis. \textbf{b)} Homemade setup based on a cylindrical Spinside dielectric cavity, rotatable around the vertical axis together with the sample. In the sample frame, $b_1$ is always aligned with the c-axis; rotating the cavity-sample assembly is equivalent to rotating $B_0$ in the ac-plane.}
		\label{fig:setups}
	\end{figure}
	
	With $g_\perp = 0$, driving terbium transitions requires a non-zero component of the microwave field $\bm{b_1}$ along the c-axis. Two approaches were used in this work. The first relies on a commercial spectrometer (Bruker with MD5 cavity) in which $\bm{b_1} \perp \bm{B_0}$, both fixed in the laboratory frame, with the sample tilted by an angle $\theta_\mathrm{mw} < \pi/2$ to obtain a non-zero c-axis projection of $\bm{b_1}$ (\cref{fig:setups}a). As a consequence, $\bm{B_0}$ cannot be fully aligned with the c-axis and the Rabi frequency at fixed input power is reduced by a factor $\cos^2\theta_\mathrm{mw}$. The second approach uses a home-built insert with a SpinSide cylindrical dielectric cavity (\SI{2}{mm} inner diameter, resonant at \SI{9.52}{GHz}, total quality factor $\approx 200$) whose $\bm{b_1}$ field can be rotated in the laboratory frame (\cref{fig:setups}b). The cavity plus sample assembly can be rotated in situ around the vertical axis, reconfiguring from parallel ($\bm{b_1} \parallel \bm{B_0}$) to perpendicular ($\bm{b_1} \perp \bm{B_0}$) mode while keeping $\bm{b_1} \parallel c$.
	
	\section{ESEEM suppression criterion}
	\label{app:eseem}
	
	\begin{figure}[!ht]
		\centering
		\includegraphics[width=8.6cm]{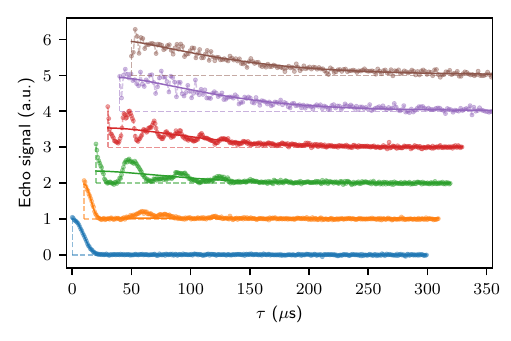}
		\caption{Measured Hahn-echo decay at \SI{6.5}{K} for different resonant fields (dots) and stretched-exponential fits (solid lines): \SI{80}{G} (blue), \SI{189}{G} (orange), \SI{334}{G} (green), \SI{589}{G} (red), \SI{1254}{G} (purple), \SI{1758}{G} (brown). Curves are offset vertically for clarity. ESEEM depth decreases and modulation frequencies change as the resonant field is increased and the nuclear Zeeman energy grows.}
		\label{fig:eseem}
	\end{figure}
	
	Electron spin echo envelope modulation (ESEEM) modulates the Hahn-echo amplitude as a function of $\tau$, owing to coherent hyperfine couplings between the probed electron spin and surrounding nuclear spins~\cite{Schweiger2001PulsedEPR,Rowan1965}. In \ce{CaWO4}, the \ce{^183W} nuclear spin produces deep multi-frequency modulations at low field that distort the echo envelope and preclude reliable stretched-exponential fits. We restrict the analysis to $B_r \geq \SI{1000}{G}$, where the modulation depth is sufficiently suppressed by the increased nuclear Zeeman energy (\cref{fig:eseem}). 
	
	\section{Measured and estimated spin-lattice relaxations}
	\label{app:SLR}
	
	\subsection{Molybdenum}
	\label{app:Mo}
	
	\begin{figure}[!ht]
		\centering
		\includegraphics[width=8.6cm]{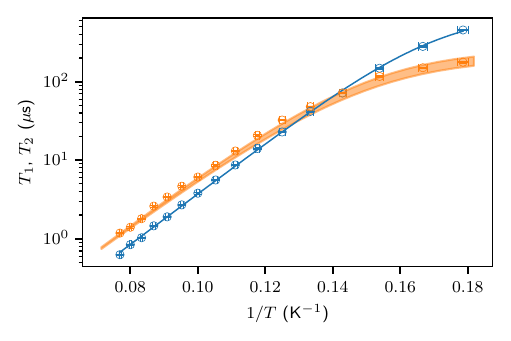}
		\caption{Blue dots, measured $T_1$ by inversion recovery of \ce{Mo^5+}; blue solid line, fit assuming direct and Orbach processes. Orange dots, measured $T_2$; shades, $T_2$ estimated from the spin-bath model.}
		\label{fig:Mo_T1}
	\end{figure}
	
	In the absence of published relaxation data for \ce{Mo^5+} in \ce{CaWO4}, we measured both its CW spectrum and its pulsed relaxation. From the CW spectrum we extract $g_{\parallel,\mathrm{Mo}} = 1.987$, $g_{\perp,\mathrm{Mo}} = 1.887$. The pulsed $T_1$ data (\cref{fig:Mo_T1}) are fitted by a direct and Orbach process,
	\begin{equation}
		R_\mathrm{Mo} = \alpha^\mathrm{Mo}_d\,T + \alpha^\mathrm{Mo}_O\,e^{-\Delta^\mathrm{Mo}_O/T},
	\end{equation}
	with $\alpha^\mathrm{Mo}_d = \SI{0.27 \pm 0.02}{(ms\cdot K)^{-1}}$, $\alpha^\mathrm{Mo}_O = \SI{0.46 \pm 0.03}{ns^{-1}}$, $\Delta^\mathrm{Mo}_O = \SI{74.5 \pm 0.6}{K}$. The measured \ce{Mo^5+} Hahn-echo coherence is limited by $T_1$ above \SI{7}{K} and by instantaneous plus spectral diffusion at lower temperature.
	
	\subsection{Dysprosium}
	\label{app:Dy}
	
	\begin{figure}[!ht]
		\centering
		\includegraphics[width=8.6cm]{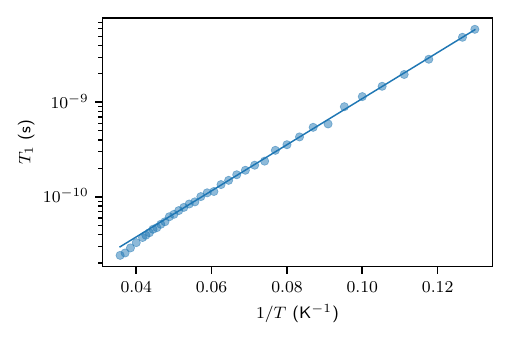}
		\caption{Dots, $T_1$ of \ce{Dy^3+} extracted from the CW linewidth; solid line, Orbach fit.}
		\label{fig:Dy_T1}
	\end{figure}
	
	From CW spectra we extract $g_{\parallel,\mathrm{Dy}} = 7.267$ and $g_{\perp,\mathrm{Dy}} = 5.466$. \ce{Dy^3+} could not be saturated in our temperature range, so an Orbach process fit to the CW-linewidth-derived $T_1$ (\cref{fig:Dy_T1}) yields
	\begin{equation}
		R_\mathrm{Dy} = \alpha^\mathrm{Dy}_O\,e^{-\Delta^\mathrm{Dy}_O/T},
	\end{equation}
	with $\alpha^\mathrm{Dy}_O = \SI{0.249 \pm 0.006}{ps^{-1}}$ and $\Delta^\mathrm{Dy}_O = \SI{56.0 \pm 0.3}{K}$. Within our temperature range the spin-flip time is below \SI{100}{ns}, placing \ce{Dy^3+} deep in the motional-narrowing regime with $T_2$ limits above \SI{500}{ms}.
	
	\subsection{Er, Ce, Nd}
	\label{app:ErCeNd}
	
	For the other Kramers ions we use literature $g$-tensors and relaxation rates. For \ce{Er^3+} ($g_{\parallel,\mathrm{Er}} = 1.247$, $g_{\perp,\mathrm{Er}} = 8.38$)~\cite{antipin_paramagnetic_1968}:
	\begin{equation}
		R_\mathrm{Er} = \alpha^\mathrm{Er}_d\,T + \alpha^\mathrm{Er}_{O,1}\,e^{-\Delta^\mathrm{Er}_{O,1}/T} + \alpha^\mathrm{Er}_{O,2}\,e^{-\Delta^\mathrm{Er}_{O,2}/T},
	\end{equation}
	with $\alpha^\mathrm{Er}_d = \SI{4.6}{s^{-1}}$, $\alpha^\mathrm{Er}_{O,1} = \SI{1.25e7}{s^{-1}}$, $\alpha^\mathrm{Er}_{O,2} = \SI{4.5e10}{s^{-1}}$, $\Delta^\mathrm{Er}_{O,1} = \SI{26}{K}$, $\Delta^\mathrm{Er}_{O,2} = \SI{58}{K}$. For \ce{Ce^3+} ($g_{\parallel,\mathrm{Ce}} = 2.92$, $g_{\perp,\mathrm{Ce}} = 1.43$)~\cite{KielMims1967}:
	\begin{equation}
		R_\mathrm{Ce} = \alpha^\mathrm{Ce}_d\,T + \alpha^\mathrm{Ce}_R\,T^{n_\mathrm{Ce}},
	\end{equation}
	with $\alpha^\mathrm{Ce}_d = \SI{0.14}{s^{-1}}$, $\alpha^\mathrm{Ce}_R = \SI{1.4e-6}{s^{-1}K^{-n_\mathrm{Ce}}}$, $n_\mathrm{Ce} = 10.9$. For \ce{Nd^3+} ($g_{\parallel,\mathrm{Nd}} = 2.03$, $g_{\perp,\mathrm{Nd}} = 2.52$)~\cite{KielMims1967}:
	\begin{equation}
		R_\mathrm{Nd} = \alpha^\mathrm{Nd}_d\,T + \alpha^\mathrm{Nd}_R\,T^{n_\mathrm{Nd}},
	\end{equation}
	with $\alpha^\mathrm{Nd}_d = \SI{2.3}{s^{-1}}$, $\alpha^\mathrm{Nd}_R = \SI{4e-6}{s^{-1}K^{-n_\mathrm{Nd}}}$, $n_\mathrm{Nd} = 10.4$.
	
	\subsection{Field dependence of impurity $T_1$}
	\label{app:T1_field}

\begin{figure}[!ht]
		\centering
		\includegraphics[width=8.6cm]{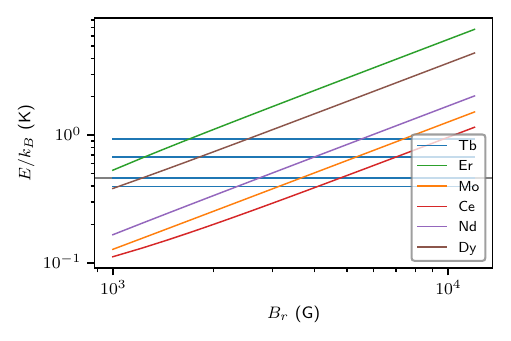}
		\caption{Computed energy transition, expressed in Kelvin, of the different ions (\ce{Tb^3+} in blue, \ce{Mo^5+} in orange, \ce{Er^3+} in green, \ce{Ce^3+} in red, \ce{Nd^3+} in purple and \ce{Dy^3+} in brown) as a function of resonant field $B_r$ of line (L3). Horizontal gray line corresponds to the resonant frequency of the X-band cavity.}
		\label{fig:Elvl_impurities}
	\end{figure}
    
	\begin{figure}[!ht]
		\centering
		\includegraphics[width=8.6cm]{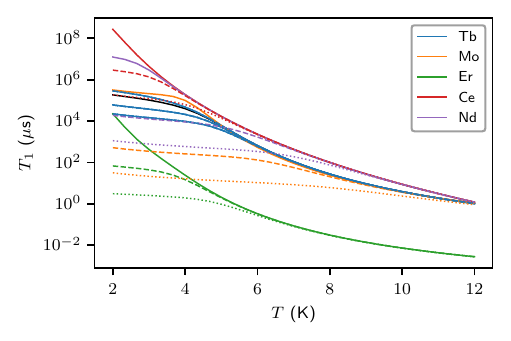}
		\caption{Estimated $T_1$ of the different impurities (probed \ce{Tb^3+} line in black, unprobed \ce{Tb^3+} line in blue, \ce{Mo^5+} in orange, \ce{Er^3+} in green, \ce{Ce^3+} in red, \ce{Nd^3+} in purple) as a function of temperature for a resonant field of the probed \ce{Tb^3+} line (L3) at \SI{1000}{G} (solid), \SI{5000}{G} (dashed) and \SI{10000}{G} (dotted).}
		\label{fig:T1_impurities}
	\end{figure}
	
	The longitudinal relaxation rates of the bath ions depend on their spin transition frequency, which in turn depends on the resonant field of the probed terbium line through the Zeeman splitting (\cref{fig:Elvl_impurities}). At fixed $T$, the spin-flip rate of a given Kramers species can therefore span orders of magnitude from $R\tau \ll 1$ at low field to $R\tau \gg 1$ at high field (\cref{fig:T1_impurities}). For the unprobed terbium lines, the transition frequency does not depend on the magnitude of the probed field and $R\tau \ll 1$ throughout the low-temperature range.
	
	\section{Model of decoherence contributions}
	\label{app:model}
	\subsection{Schematic of decoherence mechanisms}
	\label{app:schematic}
	
	\begin{figure}[!ht]
		\centering
		\includegraphics[width=\textwidth]{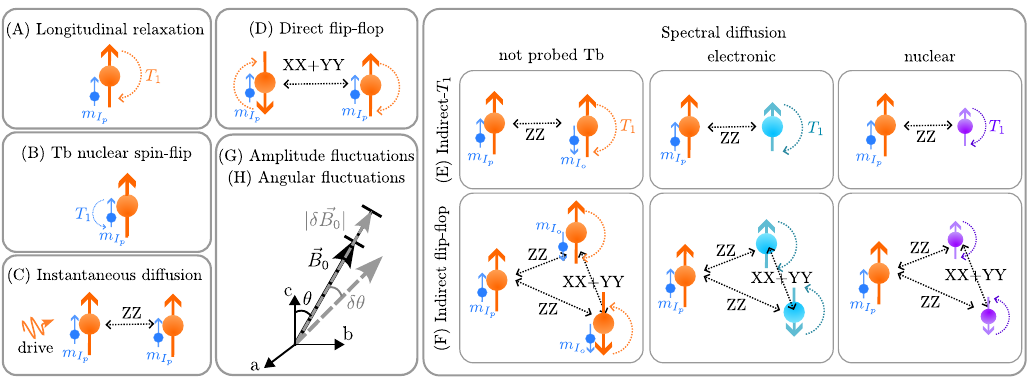}
		\caption{Schematic of decoherence mechanisms for the probed \ce{Tb^3+} spin at nuclear projection $m_{I_p}$ (orange and blue arrows). (A) spin-lattice longitudinal relaxation; (B) instantaneous diffusion; (C) terbium nuclear spin-flip; (D) direct electronic flip-flop; spectral diffusion via (E) indirect-$T_1$ and (F) indirect flip-flop from other \ce{Tb^3+} lines ($m_{I_o}$) or from bath electronic and nuclear spins; (G) external magnetic field amplitude fluctuations; (H) angular fluctuations, global (mechanical vibrations, $\delta\theta$) and local (phonon-driven modulation of the crystal axes).}
		\label{fig:drawing}
	\end{figure}

    A schematic of different decoherence mechanisms is shown in \cref{fig:drawing}. For each mechanism $a$, we compute an effective decay rate $\Gamma_\mathrm{eff, a} (\tau)$ which can depend on the inter-pulse delay $\tau$. The computation for each mechanism is described in the following sections. At the end, as we are fitting our data with a stretched-exponential decay, we compute the total effective $T_2$ by solving 
    $(\frac{2\tau}{T_2})^\beta = \sum_a \tau \Gamma_\mathrm{eff, a} (\tau)$ estimated at $\tau = T_2/2$ such that $1 = \sum_a \frac{T_2}{2} \Gamma_\mathrm{eff, a} (\frac{T_2}{2})$.

\subsection{Direct \ce{Tb^3+} flip-flops}
	\label{app:flipflop}
	
	For Kramers ions, flip-flop is driven by the transverse magnetic moment $\mu_{x,y} \propto g_\perp$. For \ce{Tb^3+}, $g_\perp = 0$ and this standard mechanism is absent. The flip-flop coupling instead arises from the tunnel-induced off-diagonal matrix element of the \emph{longitudinal} magnetic moment~\cite{Car2019,altshuler_phase_1980}:
	\begin{equation}
		\langle +|\mu_z|-\rangle = -\frac{g_\parallel\mu_B}{2}\,\frac{h\Delta}{E} = -\frac{g_\parallel\mu_B}{2}\,\sin\vartheta,
	\end{equation}
	with $E = h\sqrt{\nu_z^2 + \Delta^2}$ the transition energy, $h\nu_z = (g_\parallel\mu_B B\cos\theta + hA_\parallel m_I)$ the Zeeman splitting, and the mixing angle $\vartheta$ defined by $\sin\vartheta = h\Delta/E$. This mechanism depends on the \emph{magnitude} of the applied field, not just its direction, through the competition between $\Delta$ and $\nu_z$.
	
	Following Ref.~\cite{Car2019} adapted to the non-Kramers quasi-doublet, the flip-flop coupling factor (angularly averaged over pair orientations) is
	\begin{equation}
		\Xi_\mathrm{NK} = \frac{g_\parallel^4}{20}\,\sin^4\vartheta = \frac{g_\parallel^4}{20}\,\left(\frac{\Delta^2}{\nu_z^2 + \Delta^2}\right)^{\!2},
	\end{equation}
	with a flip-flop rate
	\begin{equation}
		R_\mathrm{FF} = \frac{\mu_0^2\mu_B^4}{12\hbar^2}\,\frac{\Xi_\mathrm{NK}\,n_s^2}{\Gamma_\mathrm{inh}},
	\end{equation}
	where $n_s$ is the spin density and $\Gamma_\mathrm{inh}$ the inhomogeneous linewidth.

    The coupling is governed by the characteristic field $B^* = h\Delta/(g_\parallel\mu_B) \approx \SI{330}{G}$. For $B \gg B^*$ along the c-axis, the Zeeman energy dominates and $\Xi_\mathrm{NK} \propto (\Delta/\nu_z)^4 \to 0$; for $B \ll B^*$ or $\theta \to \pi/2$, the tunnel splitting dominates and $\Xi_\mathrm{NK} \to g_\parallel^4/20$ is maximal.
	A consequence is that, at fixed microwave frequency $\nu_c$, the resonance condition fixes $\nu_z^2 = \nu_c^2 - \Delta^2 = \text{const.}$, so $\sin^4\vartheta$ is independent of the crystal orientation $\theta$, a qualitative departure from the Kramers case~\cite{Car2019}. 
    At resonant field in X-band, we have $\sin^4\vartheta \simeq 0.54$. 
    For $C_\mathrm{Tb} = \SI{15}{ppb}$ this yields $R_\mathrm{FF} \lesssim \SI{e-2}{s^{-1}}$, or a flip-flop lifetime above \SI{e2}{s}, which is negligible compared to the measured $T_2 \sim \SI{150}{\micro s}$ or spin-lattice relaxation $T_1 \leq \SI{0.2}{s}$ in our range of temperature.

\subsection{Dipolar linewidth for any magnetic field orientation}
\label{app:dipolar_linewidth}
We want to compute the dipolar linewidth used for the computation of instantaneous diffusion and spectral diffusion. We follow and adapt the computation from Refs. \cite{MARYASOV1982, LeDantecPhD, Lim2018}.

\subsubsection{Effective moment and quantization axis}

Each ion is assumed as an effective spin \(S = 1/2\) with axial \(g\)-tensor $\mathbf g_i = \mathrm{diag}(g_{\perp i}, g_{\perp i}, g_{\parallel i}),$
with symmetry axis \(c \parallel \hat{\mathbf z}\). The static field is
\[
\mathbf B_0 = B_0 \mathbf n, \qquad \mathbf n = (\sin\theta, 0, \cos\theta),
\]
which makes an angle \(\theta\) with \(c\).\\
Writing \(\mathbf g_i \mathbf n = g_i(\theta)\mathbf k_i\) defines the effective \(g\)-factor $g_i(\theta)$ and the quantization axis $\mathbf k_i$:
\begin{equation}
g_i(\theta) = \sqrt{g_{\parallel i}^2 \cos^2\theta + g_{\perp i}^2 \sin^2\theta},
\qquad
\mathbf k_i = \frac{1}{g_i(\theta)}\bigl(g_{\perp i}\sin\theta,\,0,\,g_{\parallel i}\cos\theta\bigr).
\end{equation}

In an eigenstate \(m_i=\pm \tfrac12\) of \(S_{k_i}\), the static magnetic moment is $$\langle \bm\mu_i\rangle = -\mu_B m_i \mathbf u_i = -\mu_B m_i \mathbf g_i \mathbf k_i = \mu_B m_i \frac{1}{g_i(\theta)}\bigl(g_{\perp i}^2 \sin\theta,\,0,\,g_{\parallel i}^2 \cos\theta\bigr)$$

The vector \(\mathbf u_i\), not \(g_i\), with $|\mathbf u_i| = \frac{\sqrt{g_{\perp i}^4 \sin^2\theta + g_{\parallel i}^4 \cos^2\theta}}{g_i(\theta)} \ge g_i(\theta)$, is what enters the dipolar coupling. The two vectors coincide only when the field lies along a principal axis. For Tb\(^{3+}\) (\(g_\perp=0\)) the moment is locked along \(c\), \(\mathbf u_{\mathrm{Tb}} = g_{\parallel}\hat{\mathbf z}\) for all \(\theta\), with \(|\mathbf u_{\mathrm{Tb}}|=g_{\parallel}\) independent of \(\theta\) even though \(g_{\mathrm{Tb}}(\theta)=g_{\parallel}\cos\theta \to 0\).

\subsubsection{Secular coupling and dipolar linewidth}

The magnetic dipole-dipole Hamiltonian, truncated to terms diagonal in both \(S_{k_a}\) and \(S_{k_b}\), by neglecting non-resonant \(a\)-\(b\) flip-flops, gives the secular coupling
\begin{equation}
H^{\mathrm{sec}}_{\mathrm{dd}} = \hbar A(\mathbf r)\, S_{k_a}S_{k_b},
\qquad
A(\mathbf r) = \frac{\mu_0\mu_B^2}{4\pi\hbar\, r^3} F({\mathbf r}),
\qquad
F({\mathbf r}) = \mathbf u_a\!\cdot\!\mathbf u_b - 3(\mathbf u_a\!\cdot\!{\mathbf e_r})(\mathbf u_b\!\cdot\!{\mathbf e_r})
\end{equation}
with ${\mathbf e_r} = {\mathbf r}/\vert {\mathbf r}\vert$.
This bilinear form is exact: it equals the diagonal matrix element of the full pair Hamiltonian in the Zeeman eigenbasis. The statistical average over a random, dilute, uncorrelated bath yields an exact Lorentzian, with FWHM
\begin{equation}
\Gamma_{\mathrm{dd}}(\theta, a, b) 
=
\frac{\mu_0\mu_B^2 C_b}{3\hbar}\,
|\mathbf u_a|\,|\mathbf u_b|\, \mathcal{J}(\psi)
\end{equation}
where $C_b$ is the concentration of spin $b$, \(\psi\) is the angle between the two moment vectors (which depends on the resonant field $B_r$, its orientation $\theta$ and the two considered spin species $a$ and $b$) and \(\mathcal{J}(\psi)\) is the universal angular integral
\begin{equation}
\mathcal{J}(\psi)
= \frac{1}{8} \int_\mathrm{sphere} \frac{\vert F({\mathbf r}) \vert }{|\mathbf u_a|\,|\mathbf u_b|} d\Omega
\qquad
\mathcal{J}(0)=\frac{2\pi}{3\sqrt3}\simeq 1.2,\qquad
\mathcal{J}\!\left(\frac{\pi}{2}\right)=1.
\end{equation}

The integral \(\mathcal{J}\) is a geometric scaling factor. It is numerically computed and varies by about 20\% over the full range. The angular dependence of \(\Gamma_{\mathrm{dd}}\) is dominated by the moduli \(|\mathbf u_i|(\theta)\). At \(\theta=0\), \(|\mathbf u_a|=g_{\parallel a}\), \(|\mathbf u_b|=g_{\parallel b}\), \(\psi=0\), and the result reduces to the standard collinear expression
\[
\Gamma_{\mathrm{dd}}(0, a, b)
=
\frac{2\pi}{9\sqrt3}\,
\frac{\mu_0\mu_B^2}{\hbar}\,
g_{\parallel a}g_{\parallel b}\,C_b.
\]

\subsubsection{Non-Kramers projection factor \(\eta\)}

For non-Kramers ions like Tb\(^{3+}\), the tunnel splitting \(\Delta\) term modify the quantization axis while the zero transverse coupling still imply that the vector $\bm{u_i}$ is along $c$. Due to the term $\Delta$, the magnitude $\abs{\bm u_i}$ is reduced by a factor 
$\eta(m_I) = \frac{\sqrt{\nu(m_I)^2-\Delta^2}}{\nu(m_I)}$ with $\nu(m_I)$ the spin transition frequency for the terbium hyperfine line $m_I$. Therefore, at the clock transition $\nu = \Delta$, there is no instantaneous or spectral diffusion on the terbium ions.  
Between two terbium ions, with moments always along c, the dipolar linewidth is independent of $\theta$ and is given by
\begin{equation}
    \Gamma_{\mathrm{dd}}(\not{\theta}, Tb(m_I), Tb(m_I'))
=
\frac{2\pi}{9\sqrt3}\,
\frac{\mu_0\mu_B^2}{\hbar}\,\eta(m_I)\eta(m_I')
g_{\parallel}^2\,C_{Tb(m_I')}.
\end{equation}
 where $\not\theta$ emphasizes the absence of dependence on orientation $\theta$.

\subsection{Instantaneous diffusion}
	\label{app:ID}
	
	Instantaneous diffusion arises from the microwave-driven flips of \ce{Tb^3+} spins resonant with the probed line during the refocusing $\pi$ pulse~\cite{Schweiger2001PulsedEPR}. The flips abruptly shift the local dipolar field on the central spin and cause irreversible dephasing at a rate
	\begin{equation}
		\frac{1}{T^\mathrm{ID}_2} = \Gamma_{dd}(\not\theta, Tb, Tb) \frac{\Gamma_{mw}}{\Gamma_{inh}(\theta)}\left\langle\sin^2(\phi_2/2)\right\rangle_f,
	\end{equation}
	where $\langle\sin^2(\phi_2/2)\rangle_f$ the average inversion probability over the EPR lineshape $f$ and the effective terbium concentration in the dipolar linewidth is reduced by the fraction within the excitation bandwidth $\Gamma_\mathrm{mw}$ when $\Gamma_\mathrm{inh} > \Gamma_\mathrm{mw}$.

	\subsection{Spectral diffusion from off-resonant \ce{Tb^3+} lines}
	\label{app:SD_Tb}
	
	The three \ce{Tb^3+} hyperfine lines with $m'_I \neq m_I$ are not excited by the microwave pulse but exert a fluctuating dipolar field on the probed spin via their longitudinal magnetic moment. Their state changes through spin-lattice relaxation (indirect-$T_1$ process); indirect flip-flops within each sub-ensemble are negligible. The spectral diffusion linewidth from off-resonant terbium with nuclear projection $m'_I$ is
	\begin{equation}
		\Gamma^{\mathrm{Tb},m'_I}_\mathrm{SD} = \Gamma_\mathrm{dd}(\not\theta, \mathrm{Tb}(m_I), \mathrm{Tb}(m'_I))\,\text{sech}^2\!\left(\frac{\Delta E_{\mathrm{Tb}(m'_I)}}{2k_B T}\right),
	\end{equation}

	Within our temperature range ($T > \SI{2}{K}$) and at the relevant fields, the hyperbolic secant factor is close to unity. The $m'_I$ dependence must be computed line by line and cannot be averaged: for $\delta m = m'_I - m_I = -1$ at X-band, $\sqrt{\nu_c^2 - \Delta^2} + A_\parallel\delta m \approx \SI{-1.4}{GHz}$, approaching a partial clock condition where the coupling to the probed spin is strongly suppressed.
	
	In the slow-flip regime $R_\mathrm{Tb}\tau \ll 1$ (relevant at low temperature), the contribution to the Hahn-echo decay rate is
	\begin{equation}
		T^\mathrm{Tb}_2 = 2~\sqrt{\frac{2}{R_{\mathrm{Tb}, m'_I}\Gamma^{\mathrm{Tb}, m'_I}_\mathrm{SD}}}.
	\end{equation}
	At $T = \SI{2.5}{K}$ and $\theta = 0$, the total off-resonant terbium spectral diffusion, summed over the three lines, yields $T^\mathrm{Tb}_2 \approx \SI{3.7}{ms}$, well above the measured values.
	
	\subsection{Spectral diffusion from Kramers impurities}
	\label{app:SD_Kramers}
	
	The Kramers ions \ce{Er^3+}, \ce{Dy^3+}, \ce{Mo^5+}, \ce{Ce^3+}, \ce{Nd^3+} contribute to spectral diffusion through their ZZ dipolar coupling to the probed terbium spin. 
 
    The spectral diffusion linewidth from a Kramers ion $K$ is
	\begin{equation}
    \Gamma^{\mathrm{K}}_\mathrm{SD} = \Gamma_\mathrm{dd}(\theta, \mathrm{Tb},m_I, \mathrm{K})\,\text{sech}^2\!\left(\frac{\Delta E_K}{2k_B T}\right),
	\end{equation}
	where $\Delta E_K = \mu_B g_\mathrm{K} B_r$ is the transition energy at the resonant field of the probed terbium. In the high-temperature limit ($k_B T \gg \Delta E_K$), relevant for most Kramers species in our conditions, the hyperbolic secant reduces to unity. However, at low temperature and high resonant field where the transition energy becomes large, the polarization factor starts to differ from unity.

    The spin-flip dynamics of each Kramers species is assessed separately at each $T$ and $B_r(\theta)$. The direct process of spin-lattice relaxation for Kramers ions depends on the transition frequency as $\propto \nu^5\coth(h\nu/2k_B T)$~\cite{AbragamBleaney1970}. Depending on the resulting $R_B\tau$, the appropriate regime of \cref{eq:SD_regimes1,eq:SD_regimes2,eq:SD_regimes3} is selected. In the parameter range explored here, \ce{Dy^3+} is always in the motional-narrowing regime, \ce{Mo^5+}, \ce{Ce^3+} and \ce{Nd^3+} remain in the slow-flip regime at moderate fields but can cross over to the intermediate regime up to the motional-narrowing regime at high $B_r$, and \ce{Er^3+} is mostly in an intermediate regime and crossing to the motional-narrowing regime at high resonant field, except at lowest temperature and resonant field where it is in the slow-spin-flip regime.
	
	\subsection{Three regimes of spectral diffusion}
	\label{app:SD}
	
	Depending on the ratio $R_B\tau$ between the bath state-change rate and the echo delay, spectral diffusion gives three different contributions~\cite{mims_phase_1968,bottger_optical_2006,HuHartmann1974}:

\begin{align}
\Gamma_{\mathrm{eff}}(\tau) = \tau \frac{\Gamma_{\mathrm{SD}}^{B} R_{B}}{2},
\qquad 
&T_2^{\mathrm{eff}} = 2\sqrt{\frac{2}{R_B \Gamma_{\mathrm{SD}}^{B}}},
\qquad 
R_B \tau \ll 1,
\label{eq:SD_regimes1}
\\[0.6em]
\Gamma_{\mathrm{eff}}(\tau) = \frac{\Gamma_{\mathrm{SD}}^{B}}{1.8},
\qquad 
&T_2^{\mathrm{eff}} = 2\frac{1.8}{\Gamma_{\mathrm{SD}}^{B}},
\qquad 
R_B \tau \sim 1,
\label{eq:SD_regimes2}
\\[0.6em]
\Gamma_{\mathrm{eff}}(\tau) = \frac{1}{\sqrt{\tau}}
\left( \frac{\Gamma_{\mathrm{SD}}^{B} \sqrt{2}}{\sqrt{\pi R_{B}}} \right),
\qquad 
&T_2^{\mathrm{eff}} = \frac{R_B \pi}{ (\Gamma_{\mathrm{SD}}^{B})^2}, 
\qquad 
R_B \tau \gg 1.
\label{eq:SD_regimes3}
\end{align}

	\subsection{Spectral diffusion from \ce{^183W} nuclear spins}
	\label{app:SD_W}
	
	The \ce{^183W} isotope ($I_W = 1/2$, natural abundance \SI{14.3}{\percent}) constitutes the intrinsic nuclear bath. 
    Since the Zeeman energy of the W nuclear spin, $\mu_N g_W B_r \ll k_B T$ throughout our parameter range, the nuclear bath is always in the high-temperature limit. The nuclear spin-flip rate is set by indirect coupling to the electron-spin bath and is typically in the hour range at our temperatures~\cite{mims_phase_1968}.
	As the \ce{^183W} nuclear spins have a relatively high concentration and are not randomly distributed, we cannot use the same computation for the dipolar linewidth as before (\cref{app:dipolar_linewidth}). For the equivalent problem on \ce{Er^3+} in a similar native-doped \ce{CaWO4} crystal, Ref.~\cite{LeDantec2021} reports an explicit cluster-correlation-expansion calculation that yields a $T_2$-limit of \SI{27}{ms}. A phenomenological scaling proposed by Kanai et al.~\cite{Kanai2022} gives
	\begin{equation}
		T^W_2 \simeq (2.7\times1.5\times 10^{18})\left(\frac{2}{g_\mathrm{eff}}\right)^{0.39}|g_W|^{-1.6}I_W^{-1.1}\,\frac{1}{c_W},
	\end{equation}
	with a prefactor 2.7 accounting for the non-random \ce{^183W} distribution in \ce{CaWO4}~\cite{LeDantecPhD}. Rescaling by the terbium effective gyromagnetic ratio yields $T^W_2 \gtrsim \SI{12}{ms}$ in our temperature range at $B_r \geq \SI{1000}{G}$, with further relief as $\theta \to \pi/2$.
	
	\subsection{Kramers flip-flop rates}
	\label{app:Kramers_ff}
	
	For completeness, the flip-flop rate between Kramers impurities of the same species is~\cite{Car2019}
	\begin{equation}
		R_\mathrm{FF} = \frac{\mu_0^2\mu_B^4}{12\hbar^2}\,\frac{\Xi_K\,n_s^2}{\Gamma_\mathrm{inh}},
	\end{equation}
	with
	\begin{align}
		\Xi_K &= \left(\frac{a - b}{2}\right)^{\!2}, \\
		a &= \tfrac{9}{80}(a_1 + a_2 + a_3), \\
		a_1 &= \left[\tfrac{2}{3}S - \tfrac{1}{2}\sin^2\Theta\,\Delta_1 + \tfrac{1}{2}\left(\cos^2\Theta - \tfrac{1}{3}\right)\cos(2\Phi)\,\Delta_2\right]^{\!2}, \\
		a_2 &= \tfrac{2}{9}\left[S - \cos(2\Phi)\,\Delta_2\right]^{\!2}, \\
		a_3 &= \tfrac{1}{3}\cos^2\Theta\,\sin^2(2\Phi)\,\Delta_2^2, \\
		b &= \tfrac{1}{8}\left[2S - \sin^2\Theta\,\Delta_1 - \sin^2\Theta\,\cos(2\Phi)\,\Delta_2\right],
	\end{align}
	with $S = g_x^2 + g_y^2$, $\Delta_1 = g_x^2 + g_y^2 - 2g_z^2$, $\Delta_2 = g_x^2 - g_y^2$, and $(\Theta, \Phi)$ the spherical coordinates of $\bm{B}$. Given the low concentrations of the native Kramers impurities, their flip-flop time exceeds \SI{1}{s} throughout our parameter range and is negligible.
	
	\subsection{Terbium nuclear spin-flips and residual field noise}
	\label{app:nuclear}
	\label{app:Bnoise}
	
	The large hyperfine coupling $A_\parallel = \SI{6.32}{GHz}$ and the expected slow nuclear dynamics mean that rare flips of the terbium nuclear spin during the Hahn echo cause irreversible loss of echo visibility on the selectively probed line, yielding $T_2 = T_{1,n}$ in this limit. We assume $T_{1,n}$ (not measured) is much larger than the electron $T_1$ and $T_2$ and neglect this contribution. Using the upper bound $T_2 = \SI{350}{ms}$ measured on phosphorus donors in enriched silicon in an identical magnet~\cite{Ross2019}, the instrumental field RMS noise satisfies $\delta B_\mathrm{rms} \lesssim \SI{0.02}{nT}$, imposing a $T_2$-limit of \SI{73}{ms} at $\theta = 0$ that is further relaxed as $\theta \to \pi/2$.
	
	\section{Stretched-exponential exponent and noise correlation time}
	\label{app:beta}
	
\begin{figure}
    \centering
    \includegraphics[width=0.5\linewidth]{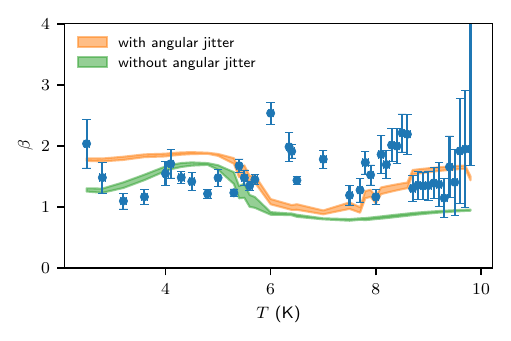}
    \caption{Stretched exponent $\beta$ as a function of temperature: dots, experimental data, shades model with (in orange) and without (in green) angular jitter.}
    \label{fig:beta_exponents2}
\end{figure}

	The stretched-exponential exponent $\beta$ extracted from $\exp[-(2\tau/T_2)^\beta]$ fits carries complementary information on the noise statistics. 
    The observed $\beta = 1.8 \pm 0.3$ in the angular-jitter-dominated regime ($T < \SI{6}{K}$, $B_r > \SI{3000}{G}$) places the relevant correlation time near $T_2$, consistent with the kHz-band attribution and a $1/f$ type of noise. 
    A more refined $(T, B_r)$ map of $\beta$ and its cross-comparison with $S_\theta(\omega)$ reconstructed by dynamical decoupling is left for future work. Indeed, we have observed some $\beta>2$ (\cref{fig:beta_exponents2}), hinting to an angular jitter noise spectral density in the form $1/f^\alpha$ with $\alpha>1$ or a more complex frequency dependence.

	\section{Decoherence contributions as a function of temperature and field}
	\label{app:Gamma_contributions}
	
	\begin{figure}[!ht]
		\centering
		\includegraphics[width=8.6cm]{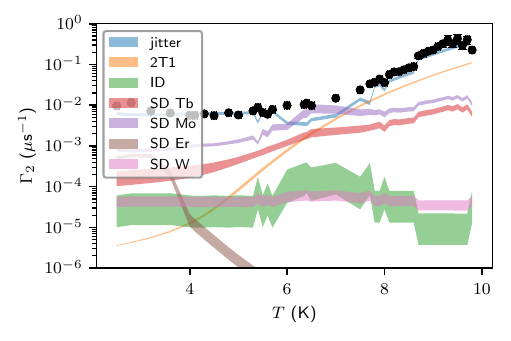}\\
		\includegraphics[width=8.6cm]{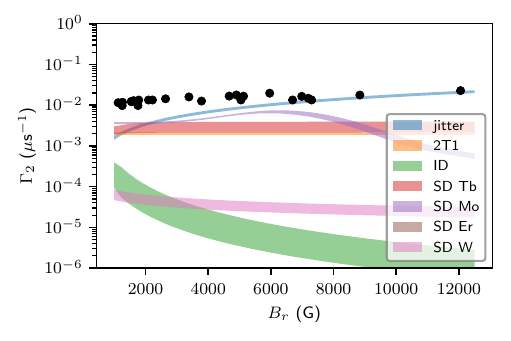}
		\caption{Decomposition of the transverse relaxation rate. Top, as a function of temperature, at the resonant field of each measurement point. Bottom, as a function of resonant field at \SI{6.5}{K}. Angular jitter is the dominant contribution across almost the entire parameter space and is the only one that grows monotonically with $B_r$.}
		\label{fig:Gamma_decomposition_vs_T}
	\end{figure}
	
	The decomposition of the transverse relaxation rate by contribution (\cref{fig:Gamma_decomposition_vs_T} top) shows that angular jitter dominates or is comparable to the largest spin-bath contribution across the full parameter range. \\
    At high temperature ($T \gtrsim \SI{8}{K}$), $1/2T_1$ and the phonon-driven part of the angular jitter are co-dominant. \\
    At low temperature ($T \lesssim \SI{6}{K}$), the global external part of angular jitter dominates, with spectral diffusion from \ce{Mo^5+} and to a lesser extent from \ce{Er^3+} and from unprobed \ce{Tb^3+} as subleading contributions. \\
    In the intermediate region, angular jitter is co-dominant with spectral diffusion from \ce{Mo^5+} and then with $1/2T_1$ and spectral diffusion from \ce{Tb^3+}.
    
    As a function of resonant field at \SI{6.5}{K} (\cref{fig:Gamma_decomposition_vs_T} bottom), most contributions are constant or decrease with resonant field except the angular jitter that always grows with $B_r$. For field below $\sim$\SI{2000}{G}, spectral diffusion from \ce{Mo^5+} is the dominant mechanism followed by angular jitter, $1/2T_1$ and spectral diffusion from \ce{Tb^3+}, above $\sim$\SI{2000}{G}, angular jitter becomes the dominant contribution.
	
	\section{Terbium $T_1$ versus orientation}
	\label{app:T1_orientation}
	
	\begin{figure}[!ht]
		\centering
		\includegraphics[width=8.6cm]{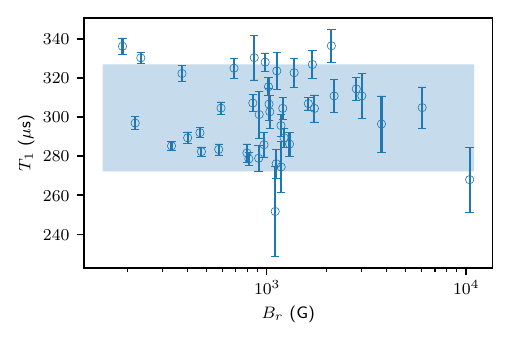}
		\caption{Measured $T_1$ for the four terbium lines as a function of resonant field $B_r$ at $T = \SI{6.4}{K}$. Shade, estimated $T_1$ envelope from \cref{eq:T1_of_T} with a \SI{0.05}{K} temperature fluctuation.}
		\label{fig:T1_vs_Br}
	\end{figure}
    The direct process for non-Kramers ions scales as $\nu^3\coth(h\nu/2k_B T)$ \cite{AbragamBleaney1970}. At fixed cavity frequency and at the resonant condition, this scaling is a constant as a function of orientation. We checked that no other mechanisms setting $T_1$ that could change as a function of orientation has been overlooked by measuring inversion-recovery $T_1$ at $T = \SI{6.4}{K}$ for the four terbium lines as a function of orientation or equivalently resonant field (\cref{fig:T1_vs_Br}). At this temperature, the Orbach process dominates. The measured $T_1$ values lie within the envelope defined by the \SI{0.05}{K} temperature stability of the cryostat. We therefore treat $T_1$, at a given transition frequency, as orientation-independent in the decoherence model of \cref{app:model}.

\section{High-temperature $T_1$-limited $T_2$}
\label{app:jitter_phonons}

\begin{figure}[!ht]
		\centering
        \includegraphics[width=8.6cm]{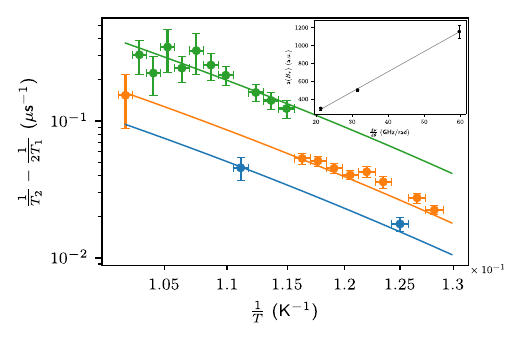}
		\caption{Measured  $1/T_2 -1/(2T_1)$ as a function of $1/T$ in the high temperature region ($T>\SI{7.5}{K}$) for three different resonant field, $B_r = \SI{1617}{G}$ in blue, $B_r = \SI{2390}{G}$ in orange and $B_r = \SI{4523}{G}$ in green. Solid lines, fits using $a(B_r)e^{-\frac{\Delta_O}{T}}$ while imposing the same Orbach gap $\Delta_O = \SI{78.8}{K}$. The fitted Orbach amplitude $a(B_r)$ depends linearly on the resonant field or equivalently on the angular sensitivity $\pdv{\nu}{\theta}$ as can be seen in the inset. The slope of $a(B_r)$ gives the coefficient $\alpha_\mathrm{ph}$ used in the main text. }
		\label{fig:jitter_phonons}
	\end{figure}

In the high temperature limit (above \SI{7.5}{K}), the $T_2$-values becomes limited by the spin-lattice longitudinal relaxation $T_1$. However, the measured $T_2(T)$ does not reach the physical limit $2T_1(T)$ and seems to scale with $a(B_r) T_1(T)$ where $a(B_r)$ does not depend on temperature but only on resonant field. Spectral diffusion may prevent reaching $2T_1$ at a given temperature but does not result in a $T_1(T)$-scaling for different temperature. Moreover, our parameter-free model including spectral diffusion cannot reproduce the dataset in the high temperature region. 

In this temperature region, $T_1$ is dominated by its Orbach process. A similar Orbach process, with an amplitude $a(B_r)$ and the same gap $\Delta_O = \SI{78.8}{K}$ fits the extra dephasing $1/T_2-1/(2T_1)$ (see \cref{fig:jitter_phonons}). The extracted $a(B_r)$ are the same order of magnitude as the Orbach amplitude $\alpha_O$ responsible for the longitudinal relaxation. This $T_1(T)$-scaling indicates another pure dephasing mechanism, probably due to an Orbach spin-phonon process as it is the mechanism setting $T_1$. For example, similarly, a spin-phonon pure dephasing contribution as it has been measured and simulated for NV centers \cite{Bar-Gill2013,Mondal2023}. 

Note that changing the resonant field does not change the terbium ions $T_1$ (\cref{app:T1_orientation}) as its transition frequencies stay constant. The only change can occur through the angular sensitivity $\pdv{\nu}{\theta}$. We model and interpret this behavior with resonant field $B_r$ in the high-temperature $T_1$-limited region as a spin-phonon pure dephasing due to phonon-induced angular jitter of the crystal axes at the unit-cell level. The field dependence discriminates this interpretation from the more obvious
alternative. A phonon modulating the crystal field modulates both $\Delta$ and the
local orientation of the axes. The first channel enters through
$\partial\nu/\partial\Delta = \Delta/\nu_c$, which is \emph{constant} along the
resonance locus, and would produce a field-independent extra dephasing. Only the
angular channel carries the $\partial\nu/\partial\theta \propto B_r\sin\theta$ scaling
that is observed in the inset of \cref{fig:jitter_phonons}. The linear growth of
$a(B_r)$ therefore selects the angular pathway.

Finally, a microscopic derivation from first principles \cite{Mondal2023} is necessary to validate our phonon-induced angular jitter and validate numerically the observed coefficient $\alpha_\mathrm{ph}$.
    
	\section{Setup-dependent angular jitter}
	\label{app:setup_jitters}
	
	\begin{figure}[!ht]
		\centering
		\includegraphics[width=8.6cm]{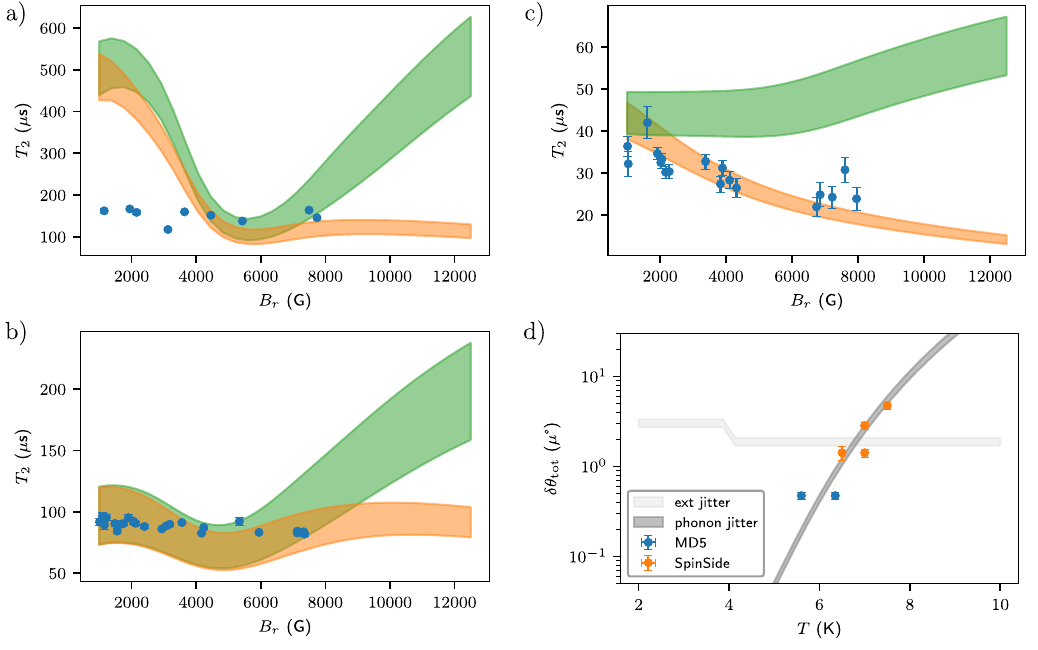}
		\caption{Measured $T_2$ as a function of resonant field (blue dots) for different setups and temperatures. Shades: model without (green) and with (orange) angular jitter. (a,b) MD5 setup at $T = \SI{5.6}{K}$ and $\SI{6.35}{K}$. (c) Spinside setup at $T = \SI{7.5}{K}$. In (d), extracted values of the total angular jitter amplitude as a function of temperature, from the $T_2$ versus $B_r$ data with MD5 setup in blue, with SpinSide setup in orange and from the $T_2$ versus temperature \cref{fig:Gamma_m}.c in gray with the contribution from external jitter in light gray and from local phonon-driven jitter in darker gray.}
		\label{fig:jitter_setups}
	\end{figure}
	
	Measuring the $B_r$ dependence of $T_2$ yields a setup- and run-specific value of the angular jitter amplitude noise $\delta\theta_\mathrm{tot} = \sqrt{\delta\theta_\mathrm{ext}^2+\delta\theta_\mathrm{ph}^2}$. The MD5 setup gives $\delta\theta_\mathrm{tot} = \SI{0.5}{\micro\degree}$ at \SI{5.6}{K} and \SI{6.35}{K}. The Spinside setup gives $\SI{1.4}{\micro\degree}$ at \SI{6.5}{K}, $\SI{1.4}{\micro\degree}$ and $\SI{2.8}{\micro\degree}$ at \SI{7}{K} depending on the tightness, and $\SI{3.8}{\micro\degree}$ at \SI{7.5}{K}. All angular jitter amplitude values are given with a \SI{10}{\percent} relative uncertainty. 
    The total angular jitter amplitude is the same order of magnitude for the different setups, \textit{i.e.}  commercial Bruker with MD5 cavity and homemade with SpinSide cavity. They are below but the same order of magnitude as the value obtained from the global fit of $T_2$ as a function of temperature, which averages over all realizations, runs and setups. 
    Trying to subtract the estimated phonon-driven contribution, that we assume to be setup independent, leads to an estimation of the external global angular jitter contribution. A positive $\delta\theta_\mathrm{ext}= \sqrt{\delta\theta_\mathrm{tot}^2-\delta\theta_\mathrm{ph}^2}$ is sometimes possible only when considering the uncertainties. It gives as an order of magnitude, $\delta\theta_\mathrm{ext} \sim \SI{0.4}{\micro\degree}$ at \SI{5.6}{K} and $\delta\theta_\mathrm{ext} \sim \SIrange{0}{0.3}{\micro\degree}$ at \SI{6.35}{K} while the Spinside setup results in $\delta\theta_\mathrm{ext} \sim \SI{1.2}{\micro\degree}$ at \SI{6.5}{K}, $\delta\theta_\mathrm{ext} \sim \SIrange{0}{0.6}{\micro\degree}$ and $\delta\theta_\mathrm{ext} \sim \SI{2.3}{\micro\degree}$ at \SI{7}{K} depending on the tightness, and $\delta\theta_\mathrm{ext} \sim \SI{1.8}{\micro\degree}$ at \SI{7.5}{K}. 
	
\end{document}